\documentclass[prl,aps,groupedaddress,superscriptaddress, twocolumn]{revtex4-2}
\usepackage{amsmath,amssymb}
\usepackage[colorlinks,bookmarks=false,citecolor=magenta,linkcolor=magenta,urlcolor=magenta]{hyperref}
\usepackage{tikz}
\usepackage{mathtools}
\usepackage{bm}
\usepackage{ragged2e}
\usepackage{graphicx}
\usepackage{subcaption}

\newcommand{\myroman}[1]{\uppercase\expandafter{\romannumeral#1}}
\usepackage[normalem]{ulem}

\begin{document}
	\title{Observation of entanglement negativity transition of pseudo-random mixed states}

	\author{Tong Liu}
	 \affiliation{Institute of Physics, Chinese Academy of Sciences, Beijing 100190, China}
	 \affiliation{School of Physical Sciences, University of Chinese Academy of Sciences, Beijing 100190, China}

	\author{Shang Liu}
	\affiliation{Kavli Institute for Theoretical Physics, University of California, Santa Barbara, California 93106, USA}

	\author{Hekang Li}
	\affiliation{Institute of Physics, Chinese Academy of Sciences, Beijing 100190, China}

	\author{Hao Li}
	\affiliation{Institute of Physics, Chinese Academy of Sciences, Beijing 100190, China}

	\author{Kaixuan Huang}
	\affiliation{Beijing Academy of Quantum Information Sciences, Beijing 100193, China}
	\affiliation{Institute of Physics, Chinese Academy of Sciences, Beijing 100190, China}

	\author{Zhongcheng Xiang}
	\affiliation{Institute of Physics, Chinese Academy of Sciences, Beijing 100190, China}
	\affiliation{School of Physical Sciences, University of Chinese Academy of Sciences, Beijing 100190, China}
	\affiliation{Hefei National Laboratory, Hefei 230088, China}
	\affiliation{Beijing Academy of Quantum Information Sciences, Beijing 100193, China}
	\affiliation{CAS Center of Excellence for Topological Quantum Computation, University of Chinese Academy of Sciences, Beijing 100190, China}
	\affiliation{Songshan Lake  Materials Laboratory, Dongguan 523808, Guangdong, China}

	\author{Xiaohui Song}
	\affiliation{Institute of Physics, Chinese Academy of Sciences, Beijing 100190, China}
	\affiliation{School of Physical Sciences, University of Chinese Academy of Sciences, Beijing 100190, China}
	\affiliation{Hefei National Laboratory, Hefei 230088, China}
	\affiliation{Beijing Academy of Quantum Information Sciences, Beijing 100193, China}
	\affiliation{CAS Center of Excellence for Topological Quantum Computation, University of Chinese Academy of Sciences, Beijing 100190, China}
	\affiliation{Songshan Lake  Materials Laboratory, Dongguan 523808, Guangdong, China}

	\author{Kai Xu}
	\email{kaixu@iphy.ac.cn}
	\affiliation{Institute of Physics, Chinese Academy of Sciences, Beijing 100190, China}
	\affiliation{School of Physical Sciences, University of Chinese Academy of Sciences, Beijing 100190, China}
	\affiliation{Hefei National Laboratory, Hefei 230088, China}
	\affiliation{Beijing Academy of Quantum Information Sciences, Beijing 100193, China}
	\affiliation{CAS Center of Excellence for Topological Quantum Computation, University of Chinese Academy of Sciences, Beijing 100190, China}
	\affiliation{Songshan Lake  Materials Laboratory, Dongguan 523808, Guangdong, China}

	\author{Dongning Zheng}
	\email{dzheng@iphy.ac.cn}
	\affiliation{Institute of Physics, Chinese Academy of Sciences, Beijing 100190, China}
	\affiliation{School of Physical Sciences, University of Chinese Academy of Sciences, Beijing 100190, China}
	\affiliation{Hefei National Laboratory, Hefei 230088, China}
	\affiliation{Beijing Academy of Quantum Information Sciences, Beijing 100193, China}
	\affiliation{CAS Center of Excellence for Topological Quantum Computation, University of Chinese Academy of Sciences, Beijing 100190, China}
	\affiliation{Songshan Lake  Materials Laboratory, Dongguan 523808, Guangdong, China}

	\author{Heng Fan}
	\email{hfan@iphy.ac.cn}
	\affiliation{Institute of Physics, Chinese Academy of Sciences, Beijing 100190, China}
	\affiliation{School of Physical Sciences, University of Chinese Academy of Sciences, Beijing 100190, China}
	\affiliation{Hefei National Laboratory, Hefei 230088, China}
	\affiliation{Beijing Academy of Quantum Information Sciences, Beijing 100193, China}
	\affiliation{CAS Center of Excellence for Topological Quantum Computation, University of Chinese Academy of Sciences, Beijing 100190, China}
	\affiliation{Songshan Lake  Materials Laboratory, Dongguan 523808, Guangdong, China}

	\begin{abstract}
		Multipartite entanglement is a key resource for quantum computation. It is expected theoretically that entanglement transition may happen for
		multipartite random quantum states, however, which is still absent experimentally.
		Here, we report the observation of entanglement transition quantified by negativity using a fully connected 20-qubit superconducting processor.
		We implement multi-layer pseudo-random circuits to generate pseudo-random pure states of 7 to 15 qubits.
		Then, we investigate negativity spectra of reduced density matrices obtained by quantum state tomography for 6 qubits.
		Three different phases can be identified by calculating logarithmic negativities based on the negativity spectra. We observe the phase transitions by changing the sizes of environment and subsystems.
		The randomness of our circuits can be also characterized by quantifying the distance between the distribution of output bit-string probabilities and Porter-Thomas distribution.
		Our simulator provides a powerful tool to generate random states and understand the entanglement structure for multipartite quantum systems.
	\end{abstract}

	\maketitle
	
	\section{Introduction}
	
	Random matrices have wide applications in various research areas.
	They may also play a significant role in quantum information processing.
	Recently, random quantum states, which are in an exponentially large computational space, are generated in demonstrating quantum supremacy for the task of sampling outputs of random quantum circuits~\cite{Arute2019,Boixo2018}.
	On the other hand, entanglement is a unique property of quantum states~\cite{RevModPhys.81.865,RevModPhys.82.277,PhysRevLett.71.666,PhysRevLett.96.110404,Calabrese2005,RevModPhys.91.021001,PhysRevLett.120.050507,PhysRevLett.93.110501,PhysRevLett.109.020505,Islam2015,Kaufman2016,Lukin2019,PhysRevLett.120.050406,Brydges2019,PhysRevA.99.052323,Huang2020}.
	It is expected that random quantum states may hold universal entanglement characteristics, which can be classified into different phases~\cite{nidari2007,PhysRevA.85.030302,PhysRevA.85.062331,Collins2016,PRXQuantum.2.030347}.

	Random quantum states can be generated by random quantum circuits with enough depth and entangling capability.
	A fully connected quantum processor may enhance the entangling power of shallow circuits compared with short-range connected processors, and facilitate the realization of random quantum states in noisy intermediate scale quantum devices (NISQ).
	Still, the measurement of entanglement is experimentally challenging.
	In this work, we report our experiments in probing entanglement negativity transition for random quantum states using a fully connected superconducting quantum processor.

	After generating multi-qubit random pure states divided into $A_1$, $A_2$ and $B$,
	we focus on the entanglement between subsystems $A_1$ and $A_2$ consisting of the system $A$, and regard $B$ as the environment.
	The entanglement is quantified by the negativity, which is a well-defined and effectively computable measure of entanglement for quantum states particularly for mixed states~\cite{PhysRevLett.77.1413,PhysRevA.65.032314,PhysRevLett.95.090503,PhysRevLett.94.040502,PhysRevLett.121.150503,PhysRevLett.125.200502,PhysRevB.94.195121,PhysRevB.101.064207,PhysRevLett.128.140502}.
	By modifying the sizes of subsystems and environment, 
	a transition from positive partial transpose (PPT) states with a vanishing negativity to negative partial transpose (NPT) states with a non-zero negativity was numerically predicted in the large Hilbert space limit~\cite{PhysRevA.85.030302,PhysRevA.85.062331}.
	Recently, another phase transition between two types of NPT states was uncovered theoretically~\cite{PRXQuantum.2.030347}.
	However, the experimental observation is still absent which is the aim of our work.

	\begin{figure*}[t]
		\centering
		\includegraphics[width=\textwidth]{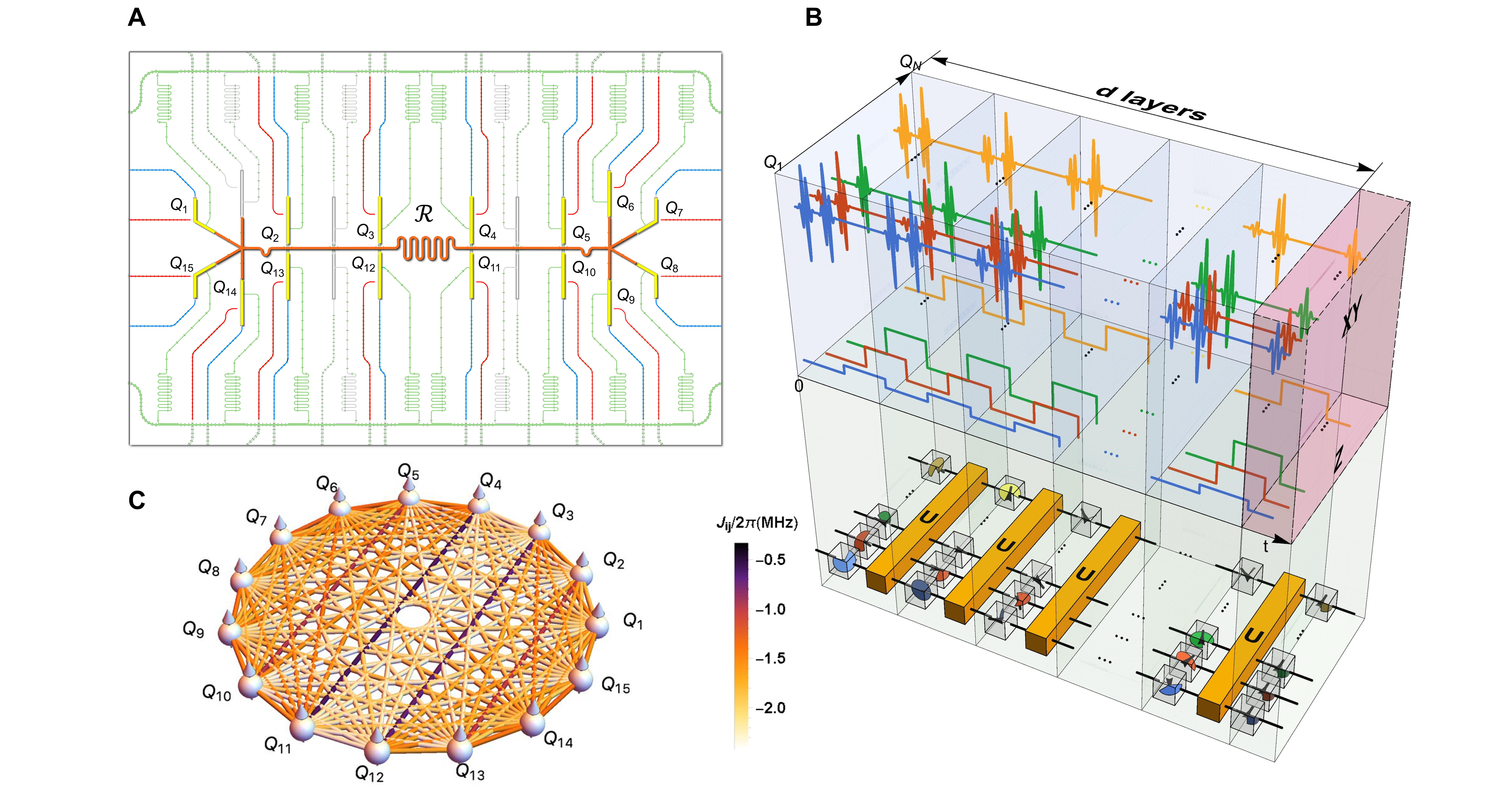}
		\captionsetup{justification=RaggedRight}
		\caption{\quad Quantum simulator and experimental pulse sequences. (A) False-color optical micrograph with highlighting various circuit elements. Qubits (yellow) are labelled from $Q_1$ to $Q_{15}$ and the central resonator (red) capacitively coupled to all qubits is labelled as $\mathcal R$. Each labelled qubit can be controlled by its XY line (red) and Z line (blue), and measured through the readout resonator (green). (B) The pulse sequences of a pseudo-random circuit and its equivalent gate model. $U$ is an entangling gate and each cubic box is a random single-qubit gate. (C) The schematic representation of the effective coupling graph of 15 qubits with an equal detuning $\Delta/2\pi \approx -360$ MHz from the central resonator.}
		\label{fig:circuit_wave}
	\end{figure*}

	There are three main considerations to accomplish the aim of the experiments.
	First, at least around a dozen qubits are needed to prepare random quantum states, such that all types of phase transitions can be observed. 
	Second, the randomness should be achieved which depends on the entangling power of circuits, and verified by, for example, Porter-Thomas distribution of the bitstring probabilities.
	These two necessities can be satisfied by using our fully connected 20-qubit programmable superconducting processor, where up to 15 qubits are used in the experiments.
	Third, multi-qubit entanglement should be generated and measured.
	Experimentally, we implement the pseudo-random quantum circuits with low decoherence errors and utilize the quantum state tomography (QST) to directly obtain reduced density matrices of 6 qubits.
	Our results represent the first experimental investigation of the entanglement negativity transition for random quantum states.
	It should be noted that much efforts have been made in studying negativity experimentally~\cite{PhysRevLett.125.200501,Neven2021}.
	Our results are based on negativity spectrum, for which density matrix of system $A$ is necessary.

	\section{Results}

	\begin{figure*}[ht]
		\centering
		\includegraphics[width=\textwidth]{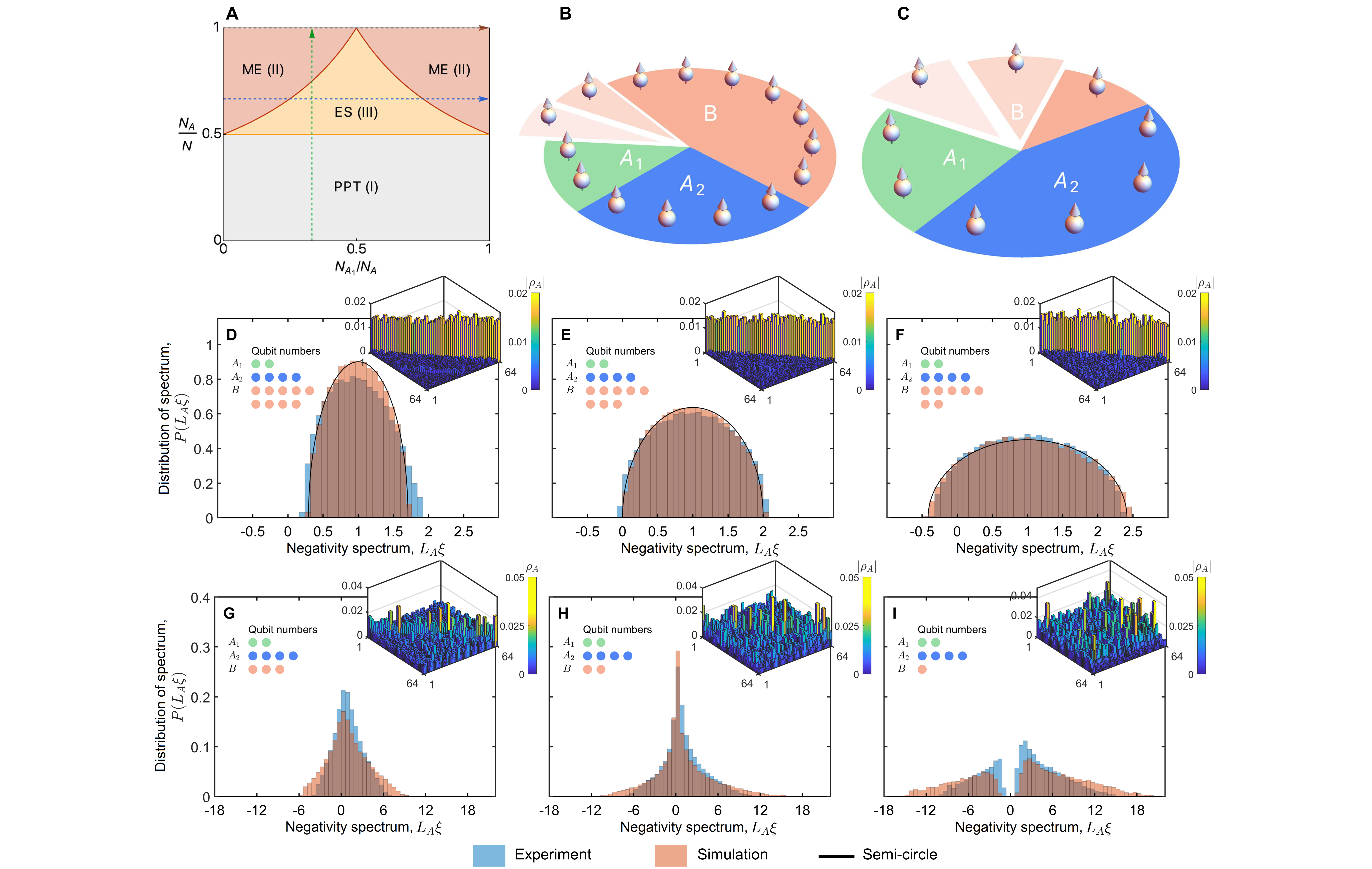}
		\captionsetup{justification=RaggedRight}
		\caption{\quad Phase diagram and negativity spetra. (A) Analytical phase diagram of reduced density matrix $\rho_A$ when $N\rightarrow\infty$. Green arrow indicates a path along which the negativity is plotted in Fig.~\ref{fig:log_neg}A. Blue or brown arrows indicate paths along which the negativity is plotted in Fig.~\ref{fig:log_neg}B. (B)-(C) Cartoons of subsystems and environment. Green, blue and orange sectors represent subsystems $A_1$, $A_2$ and environment $B$, respectively. (D)-(F) Negativity spectra sampled from pseudo-random circuits where $N_{A_1} = 2$, $N_{A_2} = 4$ and $N_B =$ 9, 8 or 7.  (G)-(I) Negativity spectra sampled from pseudo-random circuits where $N_{A_1} = 2$, $N_{A_2} = 4$ and $N_B$ = 3, 2 or 1. One of density matrices sampled from circuits with different sizes of environment is shown in the northeast corner of (D)-(I).}
		\label{fig:phase_negspec}
	\end{figure*}

    \begin{figure*}[t]
		\centering
		\includegraphics[width=\textwidth]{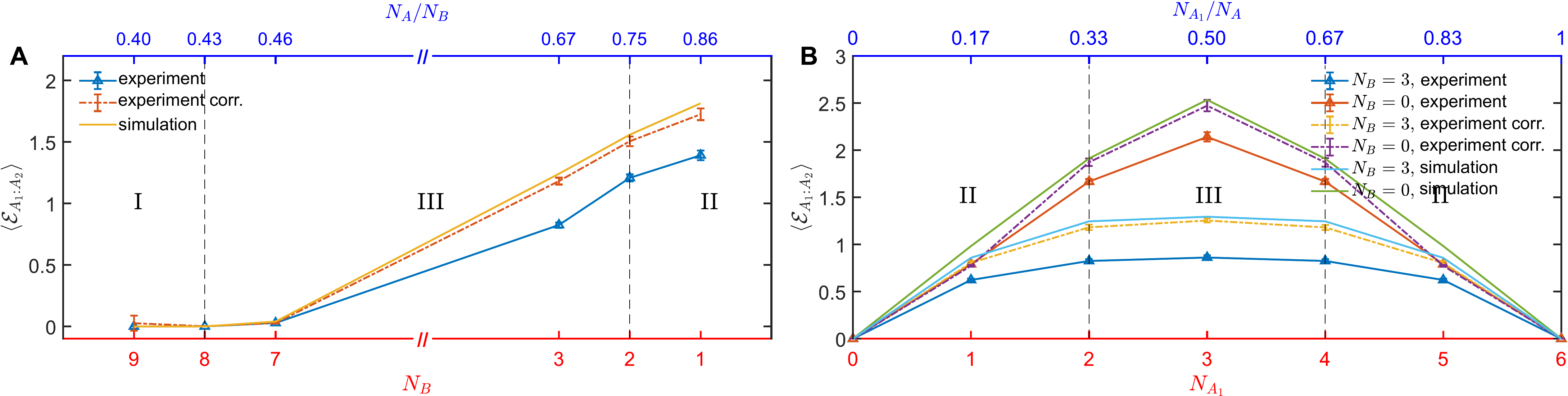}
		\captionsetup{justification=RaggedRight}
		\caption{\quad Averaged logarithmic negativity $\langle\mathcal E_{A_1:A_2}\rangle$. (A) Fix $N_{A_1} = 2$ and $N_{A_2} = 4$, then decrease $N_B$ from 9 to 1. (B) Fix $N_B$ = 3 or 0 and increase $N_{A_1}$ from $0$ to 6. When $N_B = 0$, the negativity approximately obeys volume law. In contrast, the negativity is saturated from $N_{A_1}$ = 2 to 4 when $N_B = 3$. The raw density matrices obtained from QST can be furthermore modified according to the purity~\cite{supplementary}.}
		\label{fig:log_neg}
	\end{figure*}

	Our processor contains 20 transmon qubits and one central resonator.
	All qubits are frequency-tunable and capacitively coupled to the central resonator as shown in Fig.~\ref{fig:circuit_wave}A, where qubits are labelled by $Q_j$ with $j\in\{1,2,...,15\}$ and the central resonator is denoted as $\mathcal R$.
	Each qubit can be addressed by its separate XY line and Z line, which allows us to apply one-qubit or multi-qubit gates to specified qubits.
	The implementation of a random unitary operator acting on $N$ qubits is exponentially hard~\cite{Pozniak_1998}.
	However, polynomial-sized pseudo-random quantum circuits can reproduce certain statistical properties of random ones~\cite{science.1090790,PhysRevA.76.012318,PhysRevA.78.062329,Harrow2009,PhysRevLett.104.250501,wikli_ski_2013,PhysRevLett.116.170502,PhysRevX.11.031019}.
	The other methods to generate pseudo-randomness include graph state techniques~\cite{PhysRevLett.116.200501,PhysRevA.97.022333} and random dynamics of design Hamiltonian~\cite{PhysRevX.7.021006,PhysRevLett.123.030502}.
	Considering the advantage of our processor in generation of multipartite entanglement~\cite{PhysRevLett.119.180511,Schrodingercat20,PhysRevA.100.062302}, we adopt the pseudo-random circuit approach~\cite{science.1090790} to produce pseudo-random pure states.
	A layer of pseudo-random circuit is composed of $N$ random single-qubit gates sampled from Haar measure on SU(2) group and a global entangling gate $U$.
	By controlling the amplitude and phase of Gaussian-enveloped microwave pulses transmitted by XY line, as shown in Fig.~\ref{fig:circuit_wave}B, we can fulfill different rotation gates $R_{\varphi}(\theta)$ within a 15 ns duration $\tau_\mathrm{rot}$, where $R_{\varphi}(\theta) = \exp\left[-i (\cos\varphi\sigma^x + \sin\varphi \sigma^y)\theta/2\right]$.
	In order to realize a random single-qubit gate, we decompose each single-qubit gate into two successive rotation gates $R_\varphi(\theta)$ of which rotation axes both lie in the $xy$ plane~\cite{supplementary,arXiv1303.0297S}.
	The global entangling gate $U$ acting on the $N$ qubits is defined as
	\begin{align}
		U  = \exp\left[-i\tau_{\mathrm{ent}}\sum_{i<j}^N J_{ij}(\sigma_i^+\sigma_j^- + \sigma_j^+\sigma_i^-)\right],
	\end{align}
	where $\sum_{i<j}J_{ij}(\sigma_i^+\sigma_j^-+\sigma_j^+\sigma_i^-)$ is the effective Hamiltonian of selective $N$ qubits by equally detuning them from resonator $\mathcal{R}$ with the other qubits being far off-resonant~\cite{supplementary,PhysRevLett.119.180511,Schrodingercat20}.
	$\sigma_j^{+}$ ($\sigma_j^-$) is the raising (lowering) operator for $Q_j$, $J_{ij}$ is the effective coupling strength between $Q_i$ and $Q_j$ (Fig.~\ref{fig:circuit_wave}C), and $\tau_\mathrm{ent}$ is the evolution time about 40 ns.
	As the layer of circuit increases, the measure over pseudo-random circuits converges to the Haar measure exponentially though the converge rate decreases exponentially with the number of qubits~\cite{science.1090790,PhysRevA.72.060302}.
	After applying $d$ layers of pseudo-random circuit to the initial state $|0\rangle^{\otimes N}$, we perform QST on the system qubits to estimate the reduced density matrix $\rho_A$~\cite{supplementary}.
	Then we calculate the negativity $\mathcal N_{A_1:A_2}$ between $A_1$ and $A_2$ according to the Eq.~(\ref{eq:negativity}) 
	\begin{equation}
		\mathcal N_{A_1:A_2} = \frac{\left\|\rho_A^{T_1}\right\|_1 - 1}{2},\label{eq:negativity}
	\end{equation}
	where $\|O\|_1\equiv \mathrm{Tr}(\sqrt{O^\dagger O})$ is the trace norm and $\rho_A^{T_1}$ represents the partial transpose of density matrix of $A$ with respect to subsystem $A_1$.
	$\mathcal N_{A_1:A_2}$ can also be written as the absolute value of sum of all negative eigenvalues of $\rho_A^{T_1}$
	All eigenvalues of $\rho_A^{T_1}$ constitute the negativity spectrum.
	Another associated entanglement measure, logarithmic negativity, can be induced from negativity by $\mathcal E_{A_1:A_2} = \log(2\mathcal N_{A_1:A_2} + 1)$.
	In the following context, the negativity is referred to the logarithmic negativity.

	We divide all qubits into three parts $A_1$, $A_2$ and $B$, which are comprised of $N_{A_1}$, $N_{A_2}$ and $N_B$ qubits, respectively.
	The volume or size $L_i$ of part $i$ is defined as the dimension $2^{N_i}$ of Hilbert space $\mathcal{H}_i$ where $i = A_1, A_2$ and $B$.
	$\rho_A$ is a PPT state if $L_A \equiv 2^{N_A} = 2^{N_{A_1} + N_{A_2}} > L_B/4$, otherwise it is a NPT state~\cite{PhysRevA.85.030302,PhysRevA.85.062331, PRXQuantum.2.030347}.
	NPT states can be furthermore classified into maximally entangled (ME) states and entanglement saturation states (ES) via negativity~\cite{PRXQuantum.2.030347}.

	The phase diagram of reduced density matrix $\rho_A$ (when $N\rightarrow\infty$) shown in Fig.~\ref{fig:phase_negspec}A, is divided into three phase regions PPT (\myroman{1}), ME (\myroman 2) and ES (\myroman 3) dependent on the ratio $N_{A_1}/N_A$ and $N_{A}/N$~\cite{PRXQuantum.2.030347}.
	To characterize the transition from PPT to NPT, which occurs at $N_B = N_A + 2$~\cite{PhysRevA.85.030302,PhysRevA.85.062331,PRXQuantum.2.030347}, we fix the sizes of two subsystems $N_{A_1} = 2$ and $N_{A_2} = 4$ and decrease the number of environment qubits $N_B$ from 9 to 7 by biasing the unused qubits far off-resonant, as shown in Fig.~\ref{fig:phase_negspec}B.
	After drawing 20 instances of pseudo-random circuits with five layers, of which depth is enough to capture statistical features in the simulation~\cite{supplementary}, the negativity spectra of $\rho_A$ for different environment sizes are illustrated in Fig.~\ref{fig:phase_negspec}D-F.
	The distribution of negativity spectrum is in close agreement with the semi-circle law~\cite{nidari2007,PhysRevA.85.030302,PhysRevA.85.062331, Collins2016,PRXQuantum.2.030347}, i.e.,
	\begin{equation}
		P(\xi) = \frac{2L_A}{\pi a^2}\sqrt{a^2 - \left(\xi - \frac{1}{L_A}\right)^2},\quad \left|\xi - \frac{1}{L_A}\right| < a,
	\end{equation}
	where $P(\xi)$ is the probability density of negativity spectrum and $a \equiv 2/\sqrt{L_AL_B}$ is the radius.
	Note that $N_{A_1}$, $N_{A_2}$ and $N_B$ are chosen to satisfy $L_BL_{A_2} \gg L_{A_1}$ to meet the prerequisite of semi-circle law~\cite{PhysRevA.85.062331,PRXQuantum.2.030347}.
	When $N_B = 9$, the negativity spectrum contains no negative values, which indicates that the system belongs to the PPT phase.
	Then we remove one environment qubit, sample new pseudo-random circuits and apply them to the remaining qubits.
	Now the minimum of negativity spectrum is close to zero, which shows an expected correspondence with the phase transition condition.
	Repeating the above procedure yields the negativity spectrum with 7 environment qubits, which ensures the existence of a non-zero negativity between subsystems.

	The second phase transition from ES to ME occurs at $|N_{A_1} - N_{A_2}| = N_B$.
	We still keep $N_{A_1}=2$ and $N_{A_2} = 4$, and lower $N_B$ from 3 to 1 to detect the phase transition. 
	The negativity spectra drawn from 20 instances of four-layer pseudo-random circuits are shown in Fig.~\ref{fig:phase_negspec}G-I. 
	In contrast with the negativity spectra obtained for $N_B \ge 7$,
	the negativity spectrum for $N_B = 3$ has a wider distribution and the center of the distribution is close to zero, as displayed in Fig.~\ref{fig:phase_negspec}G.
	In Fig.~\ref{fig:phase_negspec}H, we can observe a sharp peak located at zero emerging in the distribution of negativity spectrum with $N_B = 2$, which is diverging for $N\rightarrow\infty$~\cite{PRXQuantum.2.030347}.
	Next, we set $N_B = 1$ and show the distribution of negativity spectra in Fig.~\ref{fig:phase_negspec}I where we exclude some eigenvalues in the vicinity of zero~\cite{supplementary}.
	The remaining eigenvalues are split into two disjoint parts.
	The distribution of each part can be approximated by the Mar\v{c}enko-Pastur distribution~\cite{PRXQuantum.2.030347}.

	Figure~\ref{fig:log_neg}A incorporates the logarithmic negativities derived from the negativity spetra for different environment sizes.
	We start from the PPT (\myroman{1}) phase with zero negativity, enter the ES (\myroman{3}) phase at $N_B = 8$, and arrive at the ME (\myroman{2}) phase when $N_B = 2$.
	The related path in the phase diagram can be described by the vertical green line shown in Fig.~\ref{fig:phase_negspec}A.
	To distinguish ME phase and SE phase, we change the ratio between two subsystems $A_1$ and $A_2$ but hold environment invariant.
	Another benefit of QST is that we can compute the negativity between any two parts of system without additional measurements.
	As a comparison of aforementioned results, we also measure the density matrices of 6 system qubits without environment qubits.
	When $N_B = 0$, the negativity grows linearly with the number of $N_{A_1}$, which resembles a Page curve.
	However, the growth of negativity is depressed and saturated at $N_{A_1} = N_{A_2} = 3$ after taking 3 environment qubits into consideration.
	These results can be interpreted heuristically as follows: First, if there exists no environment $B$, system $A$ is totally entangled.
	Thus, the entanglement between two subsystems is proportional to the size of the minimal one.
	Second, if subsystem $A_1$ is larger than $A_2$ plus $B$, $A_2$ and $B$ will be completely entangled to $A_1$.
	Then we can deduce that there are $N_{A_2}$ pairs of entangled qubits in system $A$.
	Since the maximal entanglement pairs between $A_1$ and $A_2$ is $N_{A_2}$, we call such phase maximally entanglement phase.
	Third, if $A_1$ and $A_2$ are comparable in size and $(N_{A_1} + N_{A_2}) > N_{B}$, the environment $B$ will be entangled with $A_1$, $A_2$ in a way where $A_1$ and $A_2$ have the same number of remaining qubits to entangle with each other. 
	Hence, the entanglement between $A_1$ and $A_2$ is roughly $1/2(N_A - N_B)$ and we call such phase entanglement saturation phase.
	Finally, these results can be recapped by the following formula~\cite{PRXQuantum.2.030347}
	\begin{equation}
		\langle \mathcal E_{A_1:A_2} \rangle \approx
		\begin{cases}
		0 & N_A < N_B, \\
		\frac{1}{2}(N_A - N_B) + c_1 & N_{A_s} < \frac{N}{2} \text{ and } N_A > N_B, \\
		\text{min}(N_{A_1}, N_{A_2}) & \text{otherwise},
		\end{cases}
	\end{equation}
	where $c_1 = \log(8/3\pi)$.

	\begin{figure}[th]
		\centering
		\includegraphics[width=0.95\linewidth]{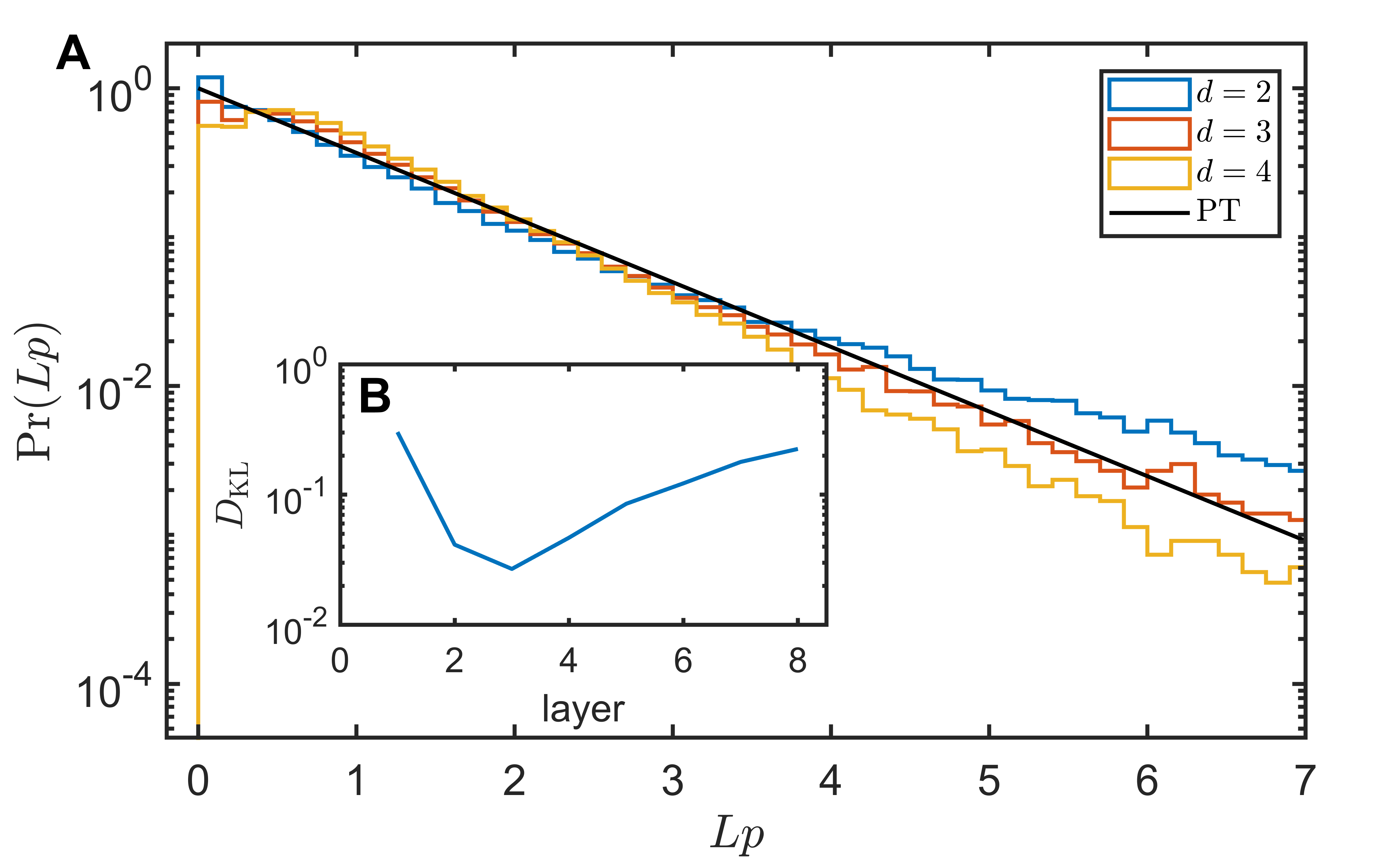}
		\captionsetup{justification=RaggedRight}
		\caption{	Histograms of output bit-string probabilities sampling from pseudo-random circuits and KL divergence. (A) Three histograms are sampled from two, three and four layers of circuits. Black sold line represents PT distribution. (B) The KL divergence between the sampling distribution and PT distribution over layers.}
		\label{fig:PT}
	\end{figure}
	Another distinctive aspect of random circuits is that the output bit-string probabilities $p(x) \equiv |\langle x|\psi\rangle|^2$ approaches the Porter-Thomas (PT) distribution, i.e., $\mathrm{Pr}(p) = Le^{-Lp}$, with increasing depth, where $|\psi\rangle$ is the output state of a circuit and $x\in\{0,1\}^N$~\cite{PhysRev.104.483,Neill2018,Boixo2018,Arute2019,PhysRevLett.127.180501}.
	Figure~\ref{fig:PT}A illustrates three histograms of the full output bit-string probabilities sampled from circuits with layers $d = 2, 3$ and 4, where small probabilities $(<1/L)$ show more often compared to the large probabilities $(>4/L)$.
	The dark solid line in Fig.~\ref{fig:PT}A represents the PT distribution.
	It is clear that the distribution from three-layer circuits is closest to the PT distribution.
	To quantify the distance between the measured distribution and PT distribution over layers, we use the Kullback-Leibler divergence defined as $D_\mathrm{KL} = S(P_\mathrm{meas}, P_\mathrm{PT}) - S(P_\mathrm{meas})$
	where $S(P_\mathrm{meas}, P_\mathrm{PT})$ is the cross entropy between the measured distribution $P_\mathrm{meas}$ and the PT distribution $P_\mathrm{PT}$, and $S(P_\mathrm{meas})$ is the self-entropy of the measured distribution~\cite{Neill2018,PhysRevA.100.062302}.
	$D_\mathrm{KL}(\ge 0)$ is zero if and only if two distributions are identical.
	As shown in Fig.~\ref{fig:PT}B, $D_\mathrm{KL}$ reaches the minimum after three layers, which verifies the observation.
	Then $D_\mathrm{KL}$ increases over layers attributed to the decoherence error~\cite{Neill2018,PhysRevA.100.062302}.

	\section{Discussion}
	We investigate the negativity spectrum of pseudo-random mixed states with the help of QST on a fully-connected quantum processor.
	The distribution of negativity spectrum can be well explained by the theories~\cite{PhysRevA.85.030302,PhysRevA.85.062331,PRXQuantum.2.030347}.
	Due to the flexibility of tuning qubits, we can observe the phase transitions by biasing specified qubits far off-resonant.
	Decoherence errors are main obstructions to precisely measure the negativity in experiments, which mix a random density matrix with a maximally mixed state and diminish the entanglement~\cite{Boixo2018,Arute2019}.
	However, major features of negativity spectrum of different phases can still be captured by our processor.
	Based on the pseudo-random circuit approach proposed in this work, our processor can be a promising platform to study the spreading of entanglement under random unitary circuits~\cite{PhysRevX.7.031016,PhysRevX.8.021014} or distinguish non-integrable and integrable systems by examining the subsystem negativity of long-time states~\cite{PhysRevB.102.235110}.

\begin{acknowledgments}
	This work was supported by National Natural Science Foundation of China (Grants Nos. 11934018, T2121001, 11904393, 92065114, 11875220 and 12047502),
	Innovation Program for Quantum Science and Technology (Grant No. 2-6),
	Strategic Priority Research Program of Chinese Academy of Sciences (Grant No. XDB28000000),
	Beijing Natural Science Foundation (Grant No. Z200009), and Scientific Instrument Developing Project of Chinese Academy of Sciences (Grant No. YJKYYQ20200041).
	S. L. is supported by the Gordon and Betty Moore Foundation under Grant No. GBMF8690 and the National Science Foundation under Grant No. NSF PHY-1748958.
\end{acknowledgments}


\begin{thebibliography}{59}%
\makeatletter
\providecommand \@ifxundefined [1]{%
 \@ifx{#1\undefined}
}%
\providecommand \@ifnum [1]{%
 \ifnum #1\expandafter \@firstoftwo
 \else \expandafter \@secondoftwo
 \fi
}%
\providecommand \@ifx [1]{%
 \ifx #1\expandafter \@firstoftwo
 \else \expandafter \@secondoftwo
 \fi
}%
\providecommand \natexlab [1]{#1}%
\providecommand \enquote  [1]{``#1''}%
\providecommand \bibnamefont  [1]{#1}%
\providecommand \bibfnamefont [1]{#1}%
\providecommand \citenamefont [1]{#1}%
\providecommand \href@noop [0]{\@secondoftwo}%
\providecommand \href [0]{\begingroup \@sanitize@url \@href}%
\providecommand \@href[1]{\@@startlink{#1}\@@href}%
\providecommand \@@href[1]{\endgroup#1\@@endlink}%
\providecommand \@sanitize@url [0]{\catcode `\\12\catcode `\$12\catcode
  `\&12\catcode `\#12\catcode `\^12\catcode `\_12\catcode `\%12\relax}%
\providecommand \@@startlink[1]{}%
\providecommand \@@endlink[0]{}%
\providecommand \url  [0]{\begingroup\@sanitize@url \@url }%
\providecommand \@url [1]{\endgroup\@href {#1}{\urlprefix }}%
\providecommand \urlprefix  [0]{URL }%
\providecommand \Eprint [0]{\href }%
\providecommand \doibase [0]{https://doi.org/}%
\providecommand \selectlanguage [0]{\@gobble}%
\providecommand \bibinfo  [0]{\@secondoftwo}%
\providecommand \bibfield  [0]{\@secondoftwo}%
\providecommand \translation [1]{[#1]}%
\providecommand \BibitemOpen [0]{}%
\providecommand \bibitemStop [0]{}%
\providecommand \bibitemNoStop [0]{.\EOS\space}%
\providecommand \EOS [0]{\spacefactor3000\relax}%
\providecommand \BibitemShut  [1]{\csname bibitem#1\endcsname}%
\let\auto@bib@innerbib\@empty
\bibitem [{\citenamefont {Arute}\ \emph {et~al.}(2019)\citenamefont {Arute},
  \citenamefont {Arya}, \citenamefont {Babbush}, \citenamefont {Bacon},
  \citenamefont {Bardin}, \citenamefont {Barends}, \citenamefont {Biswas},
  \citenamefont {Boixo}, \citenamefont {Brandao}, \citenamefont {Buell} \emph
  {et~al.}}]{Arute2019}%
  \BibitemOpen
  \bibfield  {author} {\bibinfo {author} {\bibfnamefont {F.}~\bibnamefont
  {Arute}}, \bibinfo {author} {\bibfnamefont {K.}~\bibnamefont {Arya}},
  \bibinfo {author} {\bibfnamefont {R.}~\bibnamefont {Babbush}}, \bibinfo
  {author} {\bibfnamefont {D.}~\bibnamefont {Bacon}}, \bibinfo {author}
  {\bibfnamefont {J.~C.}\ \bibnamefont {Bardin}}, \bibinfo {author}
  {\bibfnamefont {R.}~\bibnamefont {Barends}}, \bibinfo {author} {\bibfnamefont
  {R.}~\bibnamefont {Biswas}}, \bibinfo {author} {\bibfnamefont
  {S.}~\bibnamefont {Boixo}}, \bibinfo {author} {\bibfnamefont {F.~G. S.~L.}\
  \bibnamefont {Brandao}}, \bibinfo {author} {\bibfnamefont {D.~A.}\
  \bibnamefont {Buell}}, \emph {et~al.},\ }\bibfield  {title} {\bibinfo {title}
  {Quantum supremacy using a programmable superconducting processor},\ }\href
  {https://doi.org/10.1038/s41586-019-1666-5} {\bibfield  {journal} {\bibinfo
  {journal} {Nature}\ }\textbf {\bibinfo {volume} {574}},\ \bibinfo {pages}
  {505} (\bibinfo {year} {2019})}\BibitemShut {NoStop}%
\bibitem [{\citenamefont {Boixo}\ \emph {et~al.}(2018)\citenamefont {Boixo},
  \citenamefont {Isakov}, \citenamefont {Smelyanskiy}, \citenamefont {Babbush},
  \citenamefont {Ding}, \citenamefont {Jiang}, \citenamefont {Bremner},
  \citenamefont {Martinis},\ and\ \citenamefont {Neven}}]{Boixo2018}%
  \BibitemOpen
  \bibfield  {author} {\bibinfo {author} {\bibfnamefont {S.}~\bibnamefont
  {Boixo}}, \bibinfo {author} {\bibfnamefont {S.~V.}\ \bibnamefont {Isakov}},
  \bibinfo {author} {\bibfnamefont {V.~N.}\ \bibnamefont {Smelyanskiy}},
  \bibinfo {author} {\bibfnamefont {R.}~\bibnamefont {Babbush}}, \bibinfo
  {author} {\bibfnamefont {N.}~\bibnamefont {Ding}}, \bibinfo {author}
  {\bibfnamefont {Z.}~\bibnamefont {Jiang}}, \bibinfo {author} {\bibfnamefont
  {M.~J.}\ \bibnamefont {Bremner}}, \bibinfo {author} {\bibfnamefont {J.~M.}\
  \bibnamefont {Martinis}},\ and\ \bibinfo {author} {\bibfnamefont
  {H.}~\bibnamefont {Neven}},\ }\bibfield  {title} {\bibinfo {title}
  {Characterizing quantum supremacy in near-term devices},\ }\href
  {https://doi.org/10.1038/s41567-018-0124-x} {\bibfield  {journal} {\bibinfo
  {journal} {Nat. Phys.}\ }\textbf {\bibinfo {volume} {14}},\ \bibinfo {pages}
  {595} (\bibinfo {year} {2018})}\BibitemShut {NoStop}%
\bibitem [{\citenamefont {Horodecki}\ \emph {et~al.}(2009)\citenamefont
  {Horodecki}, \citenamefont {Horodecki}, \citenamefont {Horodecki},\ and\
  \citenamefont {Horodecki}}]{RevModPhys.81.865}%
  \BibitemOpen
  \bibfield  {author} {\bibinfo {author} {\bibfnamefont {R.}~\bibnamefont
  {Horodecki}}, \bibinfo {author} {\bibfnamefont {P.}~\bibnamefont
  {Horodecki}}, \bibinfo {author} {\bibfnamefont {M.}~\bibnamefont
  {Horodecki}},\ and\ \bibinfo {author} {\bibfnamefont {K.}~\bibnamefont
  {Horodecki}},\ }\bibfield  {title} {\bibinfo {title} {Quantum entanglement},\
  }\href {https://doi.org/10.1103/RevModPhys.81.865} {\bibfield  {journal}
  {\bibinfo  {journal} {Rev. Mod. Phys.}\ }\textbf {\bibinfo {volume} {81}},\
  \bibinfo {pages} {865} (\bibinfo {year} {2009})}\BibitemShut {NoStop}%
\bibitem [{\citenamefont {Eisert}\ \emph {et~al.}(2010)\citenamefont {Eisert},
  \citenamefont {Cramer},\ and\ \citenamefont {Plenio}}]{RevModPhys.82.277}%
  \BibitemOpen
  \bibfield  {author} {\bibinfo {author} {\bibfnamefont {J.}~\bibnamefont
  {Eisert}}, \bibinfo {author} {\bibfnamefont {M.}~\bibnamefont {Cramer}},\
  and\ \bibinfo {author} {\bibfnamefont {M.~B.}\ \bibnamefont {Plenio}},\
  }\bibfield  {title} {\bibinfo {title} {Colloquium: Area laws for the
  entanglement entropy},\ }\href {https://doi.org/10.1103/RevModPhys.82.277}
  {\bibfield  {journal} {\bibinfo  {journal} {Rev. Mod. Phys.}\ }\textbf
  {\bibinfo {volume} {82}},\ \bibinfo {pages} {277} (\bibinfo {year}
  {2010})}\BibitemShut {NoStop}%
\bibitem [{\citenamefont {Srednicki}(1993)}]{PhysRevLett.71.666}%
  \BibitemOpen
  \bibfield  {author} {\bibinfo {author} {\bibfnamefont {M.}~\bibnamefont
  {Srednicki}},\ }\bibfield  {title} {\bibinfo {title} {Entropy and area},\
  }\href {https://doi.org/10.1103/PhysRevLett.71.666} {\bibfield  {journal}
  {\bibinfo  {journal} {Phys. Rev. Lett.}\ }\textbf {\bibinfo {volume} {71}},\
  \bibinfo {pages} {666} (\bibinfo {year} {1993})}\BibitemShut {NoStop}%
\bibitem [{\citenamefont {Kitaev}\ and\ \citenamefont
  {Preskill}(2006)}]{PhysRevLett.96.110404}%
  \BibitemOpen
  \bibfield  {author} {\bibinfo {author} {\bibfnamefont {A.}~\bibnamefont
  {Kitaev}}\ and\ \bibinfo {author} {\bibfnamefont {J.}~\bibnamefont
  {Preskill}},\ }\bibfield  {title} {\bibinfo {title} {Topological entanglement
  entropy},\ }\href {https://doi.org/10.1103/PhysRevLett.96.110404} {\bibfield
  {journal} {\bibinfo  {journal} {Phys. Rev. Lett.}\ }\textbf {\bibinfo
  {volume} {96}},\ \bibinfo {pages} {110404} (\bibinfo {year}
  {2006})}\BibitemShut {NoStop}%
\bibitem [{\citenamefont {Calabrese}\ and\ \citenamefont
  {Cardy}(2005)}]{Calabrese2005}%
  \BibitemOpen
  \bibfield  {author} {\bibinfo {author} {\bibfnamefont {P.}~\bibnamefont
  {Calabrese}}\ and\ \bibinfo {author} {\bibfnamefont {J.}~\bibnamefont
  {Cardy}},\ }\bibfield  {title} {\bibinfo {title} {Evolution of entanglement
  entropy in one-dimensional systems},\ }\href
  {https://doi.org/10.1088/1742-5468/2005/04/p04010} {\bibfield  {journal}
  {\bibinfo  {journal} {J. Stat. Mech.}\ }\textbf {\bibinfo {volume} {2005}},\
  \bibinfo {pages} {P04010} (\bibinfo {year} {2005})}\BibitemShut {NoStop}%
\bibitem [{\citenamefont {Abanin}\ \emph {et~al.}(2019)\citenamefont {Abanin},
  \citenamefont {Altman}, \citenamefont {Bloch},\ and\ \citenamefont
  {Serbyn}}]{RevModPhys.91.021001}%
  \BibitemOpen
  \bibfield  {author} {\bibinfo {author} {\bibfnamefont {D.~A.}\ \bibnamefont
  {Abanin}}, \bibinfo {author} {\bibfnamefont {E.}~\bibnamefont {Altman}},
  \bibinfo {author} {\bibfnamefont {I.}~\bibnamefont {Bloch}},\ and\ \bibinfo
  {author} {\bibfnamefont {M.}~\bibnamefont {Serbyn}},\ }\bibfield  {title}
  {\bibinfo {title} {Colloquium: Many-body localization, thermalization, and
  entanglement},\ }\href {https://doi.org/10.1103/RevModPhys.91.021001}
  {\bibfield  {journal} {\bibinfo  {journal} {Rev. Mod. Phys.}\ }\textbf
  {\bibinfo {volume} {91}},\ \bibinfo {pages} {021001} (\bibinfo {year}
  {2019})}\BibitemShut {NoStop}%
\bibitem [{\citenamefont {Xu}\ \emph {et~al.}(2018)\citenamefont {Xu},
  \citenamefont {Chen}, \citenamefont {Zeng}, \citenamefont {Zhang},
  \citenamefont {Song}, \citenamefont {Liu}, \citenamefont {Guo}, \citenamefont
  {Zhang}, \citenamefont {Xu}, \citenamefont {Deng}, \citenamefont {Huang},
  \citenamefont {Wang}, \citenamefont {Zhu}, \citenamefont {Zheng},\ and\
  \citenamefont {Fan}}]{PhysRevLett.120.050507}%
  \BibitemOpen
  \bibfield  {author} {\bibinfo {author} {\bibfnamefont {K.}~\bibnamefont
  {Xu}}, \bibinfo {author} {\bibfnamefont {J.-J.}\ \bibnamefont {Chen}},
  \bibinfo {author} {\bibfnamefont {Y.}~\bibnamefont {Zeng}}, \bibinfo {author}
  {\bibfnamefont {Y.-R.}\ \bibnamefont {Zhang}}, \bibinfo {author}
  {\bibfnamefont {C.}~\bibnamefont {Song}}, \bibinfo {author} {\bibfnamefont
  {W.}~\bibnamefont {Liu}}, \bibinfo {author} {\bibfnamefont {Q.}~\bibnamefont
  {Guo}}, \bibinfo {author} {\bibfnamefont {P.}~\bibnamefont {Zhang}}, \bibinfo
  {author} {\bibfnamefont {D.}~\bibnamefont {Xu}}, \bibinfo {author}
  {\bibfnamefont {H.}~\bibnamefont {Deng}}, \bibinfo {author} {\bibfnamefont
  {K.}~\bibnamefont {Huang}}, \bibinfo {author} {\bibfnamefont
  {H.}~\bibnamefont {Wang}}, \bibinfo {author} {\bibfnamefont {X.}~\bibnamefont
  {Zhu}}, \bibinfo {author} {\bibfnamefont {D.}~\bibnamefont {Zheng}},\ and\
  \bibinfo {author} {\bibfnamefont {H.}~\bibnamefont {Fan}},\ }\bibfield
  {title} {\bibinfo {title} {Emulating many-body localization with a
  superconducting quantum processor},\ }\href
  {https://doi.org/10.1103/PhysRevLett.120.050507} {\bibfield  {journal}
  {\bibinfo  {journal} {Phys. Rev. Lett.}\ }\textbf {\bibinfo {volume} {120}},\
  \bibinfo {pages} {050507} (\bibinfo {year} {2018})}\BibitemShut {NoStop}%
\bibitem [{\citenamefont {Moura~Alves}\ and\ \citenamefont
  {Jaksch}(2004)}]{PhysRevLett.93.110501}%
  \BibitemOpen
  \bibfield  {author} {\bibinfo {author} {\bibfnamefont {C.}~\bibnamefont
  {Moura~Alves}}\ and\ \bibinfo {author} {\bibfnamefont {D.}~\bibnamefont
  {Jaksch}},\ }\bibfield  {title} {\bibinfo {title} {Multipartite entanglement
  detection in bosons},\ }\href {https://doi.org/10.1103/PhysRevLett.93.110501}
  {\bibfield  {journal} {\bibinfo  {journal} {Phys. Rev. Lett.}\ }\textbf
  {\bibinfo {volume} {93}},\ \bibinfo {pages} {110501} (\bibinfo {year}
  {2004})}\BibitemShut {NoStop}%
\bibitem [{\citenamefont {Daley}\ \emph {et~al.}(2012)\citenamefont {Daley},
  \citenamefont {Pichler}, \citenamefont {Schachenmayer},\ and\ \citenamefont
  {Zoller}}]{PhysRevLett.109.020505}%
  \BibitemOpen
  \bibfield  {author} {\bibinfo {author} {\bibfnamefont {A.~J.}\ \bibnamefont
  {Daley}}, \bibinfo {author} {\bibfnamefont {H.}~\bibnamefont {Pichler}},
  \bibinfo {author} {\bibfnamefont {J.}~\bibnamefont {Schachenmayer}},\ and\
  \bibinfo {author} {\bibfnamefont {P.}~\bibnamefont {Zoller}},\ }\bibfield
  {title} {\bibinfo {title} {Measuring entanglement growth in quench dynamics
  of bosons in an optical lattice},\ }\href
  {https://doi.org/10.1103/PhysRevLett.109.020505} {\bibfield  {journal}
  {\bibinfo  {journal} {Phys. Rev. Lett.}\ }\textbf {\bibinfo {volume} {109}},\
  \bibinfo {pages} {020505} (\bibinfo {year} {2012})}\BibitemShut {NoStop}%
\bibitem [{\citenamefont {Islam}\ \emph {et~al.}(2015)\citenamefont {Islam},
  \citenamefont {Ma}, \citenamefont {Preiss}, \citenamefont {Tai},
  \citenamefont {Lukin}, \citenamefont {Rispoli},\ and\ \citenamefont
  {Greiner}}]{Islam2015}%
  \BibitemOpen
  \bibfield  {author} {\bibinfo {author} {\bibfnamefont {R.}~\bibnamefont
  {Islam}}, \bibinfo {author} {\bibfnamefont {R.}~\bibnamefont {Ma}}, \bibinfo
  {author} {\bibfnamefont {P.~M.}\ \bibnamefont {Preiss}}, \bibinfo {author}
  {\bibfnamefont {M.~E.}\ \bibnamefont {Tai}}, \bibinfo {author} {\bibfnamefont
  {A.}~\bibnamefont {Lukin}}, \bibinfo {author} {\bibfnamefont
  {M.}~\bibnamefont {Rispoli}},\ and\ \bibinfo {author} {\bibfnamefont
  {M.}~\bibnamefont {Greiner}},\ }\bibfield  {title} {\bibinfo {title}
  {Measuring entanglement entropy in a quantum many-body system},\ }\href
  {https://doi.org/10.1038/nature15750} {\bibfield  {journal} {\bibinfo
  {journal} {Nature}\ }\textbf {\bibinfo {volume} {528}},\ \bibinfo {pages}
  {77} (\bibinfo {year} {2015})}\BibitemShut {NoStop}%
\bibitem [{\citenamefont {Kaufman}\ \emph {et~al.}(2016)\citenamefont
  {Kaufman}, \citenamefont {Tai}, \citenamefont {Lukin}, \citenamefont
  {Rispoli}, \citenamefont {Schittko}, \citenamefont {Preiss},\ and\
  \citenamefont {Greiner}}]{Kaufman2016}%
  \BibitemOpen
  \bibfield  {author} {\bibinfo {author} {\bibfnamefont {A.~M.}\ \bibnamefont
  {Kaufman}}, \bibinfo {author} {\bibfnamefont {M.~E.}\ \bibnamefont {Tai}},
  \bibinfo {author} {\bibfnamefont {A.}~\bibnamefont {Lukin}}, \bibinfo
  {author} {\bibfnamefont {M.}~\bibnamefont {Rispoli}}, \bibinfo {author}
  {\bibfnamefont {R.}~\bibnamefont {Schittko}}, \bibinfo {author}
  {\bibfnamefont {P.~M.}\ \bibnamefont {Preiss}},\ and\ \bibinfo {author}
  {\bibfnamefont {M.}~\bibnamefont {Greiner}},\ }\bibfield  {title} {\bibinfo
  {title} {Quantum thermalization through entanglement in an isolated many-body
  system},\ }\href {https://doi.org/10.1126/science.aaf6725} {\bibfield
  {journal} {\bibinfo  {journal} {Science}\ }\textbf {\bibinfo {volume}
  {353}},\ \bibinfo {pages} {794} (\bibinfo {year} {2016})}\BibitemShut
  {NoStop}%
\bibitem [{\citenamefont {Lukin}\ \emph {et~al.}(2019)\citenamefont {Lukin},
  \citenamefont {Rispoli}, \citenamefont {Schittko}, \citenamefont {Tai},
  \citenamefont {Kaufman}, \citenamefont {Choi}, \citenamefont {Khemani},
  \citenamefont {L{\'{e}}onard},\ and\ \citenamefont {Greiner}}]{Lukin2019}%
  \BibitemOpen
  \bibfield  {author} {\bibinfo {author} {\bibfnamefont {A.}~\bibnamefont
  {Lukin}}, \bibinfo {author} {\bibfnamefont {M.}~\bibnamefont {Rispoli}},
  \bibinfo {author} {\bibfnamefont {R.}~\bibnamefont {Schittko}}, \bibinfo
  {author} {\bibfnamefont {M.~E.}\ \bibnamefont {Tai}}, \bibinfo {author}
  {\bibfnamefont {A.~M.}\ \bibnamefont {Kaufman}}, \bibinfo {author}
  {\bibfnamefont {S.}~\bibnamefont {Choi}}, \bibinfo {author} {\bibfnamefont
  {V.}~\bibnamefont {Khemani}}, \bibinfo {author} {\bibfnamefont
  {J.}~\bibnamefont {L{\'{e}}onard}},\ and\ \bibinfo {author} {\bibfnamefont
  {M.}~\bibnamefont {Greiner}},\ }\bibfield  {title} {\bibinfo {title} {Probing
  entanglement in a many-body{\textendash}localized system},\ }\href
  {https://doi.org/10.1126/science.aau0818} {\bibfield  {journal} {\bibinfo
  {journal} {Science}\ }\textbf {\bibinfo {volume} {364}},\ \bibinfo {pages}
  {256} (\bibinfo {year} {2019})}\BibitemShut {NoStop}%
\bibitem [{\citenamefont {Elben}\ \emph {et~al.}(2018)\citenamefont {Elben},
  \citenamefont {Vermersch}, \citenamefont {Dalmonte}, \citenamefont {Cirac},\
  and\ \citenamefont {Zoller}}]{PhysRevLett.120.050406}%
  \BibitemOpen
  \bibfield  {author} {\bibinfo {author} {\bibfnamefont {A.}~\bibnamefont
  {Elben}}, \bibinfo {author} {\bibfnamefont {B.}~\bibnamefont {Vermersch}},
  \bibinfo {author} {\bibfnamefont {M.}~\bibnamefont {Dalmonte}}, \bibinfo
  {author} {\bibfnamefont {J.~I.}\ \bibnamefont {Cirac}},\ and\ \bibinfo
  {author} {\bibfnamefont {P.}~\bibnamefont {Zoller}},\ }\bibfield  {title}
  {\bibinfo {title} {R\'enyi entropies from random quenches in atomic {Hubbard}
  and spin models},\ }\href {https://doi.org/10.1103/PhysRevLett.120.050406}
  {\bibfield  {journal} {\bibinfo  {journal} {Phys. Rev. Lett.}\ }\textbf
  {\bibinfo {volume} {120}},\ \bibinfo {pages} {050406} (\bibinfo {year}
  {2018})}\BibitemShut {NoStop}%
\bibitem [{\citenamefont {Brydges}\ \emph {et~al.}(2019)\citenamefont
  {Brydges}, \citenamefont {Elben}, \citenamefont {Jurcevic}, \citenamefont
  {Vermersch}, \citenamefont {Maier}, \citenamefont {Lanyon}, \citenamefont
  {Zoller}, \citenamefont {Blatt},\ and\ \citenamefont {Roos}}]{Brydges2019}%
  \BibitemOpen
  \bibfield  {author} {\bibinfo {author} {\bibfnamefont {T.}~\bibnamefont
  {Brydges}}, \bibinfo {author} {\bibfnamefont {A.}~\bibnamefont {Elben}},
  \bibinfo {author} {\bibfnamefont {P.}~\bibnamefont {Jurcevic}}, \bibinfo
  {author} {\bibfnamefont {B.}~\bibnamefont {Vermersch}}, \bibinfo {author}
  {\bibfnamefont {C.}~\bibnamefont {Maier}}, \bibinfo {author} {\bibfnamefont
  {B.~P.}\ \bibnamefont {Lanyon}}, \bibinfo {author} {\bibfnamefont
  {P.}~\bibnamefont {Zoller}}, \bibinfo {author} {\bibfnamefont
  {R.}~\bibnamefont {Blatt}},\ and\ \bibinfo {author} {\bibfnamefont {C.~F.}\
  \bibnamefont {Roos}},\ }\bibfield  {title} {\bibinfo {title} {Probing
  {R{\'{e}}nyi} entanglement entropy via randomized measurements},\ }\href
  {https://doi.org/10.1126/science.aau4963} {\bibfield  {journal} {\bibinfo
  {journal} {Science}\ }\textbf {\bibinfo {volume} {364}},\ \bibinfo {pages}
  {260} (\bibinfo {year} {2019})}\BibitemShut {NoStop}%
\bibitem [{\citenamefont {Elben}\ \emph {et~al.}(2019)\citenamefont {Elben},
  \citenamefont {Vermersch}, \citenamefont {Roos},\ and\ \citenamefont
  {Zoller}}]{PhysRevA.99.052323}%
  \BibitemOpen
  \bibfield  {author} {\bibinfo {author} {\bibfnamefont {A.}~\bibnamefont
  {Elben}}, \bibinfo {author} {\bibfnamefont {B.}~\bibnamefont {Vermersch}},
  \bibinfo {author} {\bibfnamefont {C.~F.}\ \bibnamefont {Roos}},\ and\
  \bibinfo {author} {\bibfnamefont {P.}~\bibnamefont {Zoller}},\ }\bibfield
  {title} {\bibinfo {title} {Statistical correlations between locally
  randomized measurements: A toolbox for probing entanglement in many-body
  quantum states},\ }\href {https://doi.org/10.1103/PhysRevA.99.052323}
  {\bibfield  {journal} {\bibinfo  {journal} {Phys. Rev. A}\ }\textbf {\bibinfo
  {volume} {99}},\ \bibinfo {pages} {052323} (\bibinfo {year}
  {2019})}\BibitemShut {NoStop}%
\bibitem [{\citenamefont {Huang}\ \emph {et~al.}(2020)\citenamefont {Huang},
  \citenamefont {Kueng},\ and\ \citenamefont {Preskill}}]{Huang2020}%
  \BibitemOpen
  \bibfield  {author} {\bibinfo {author} {\bibfnamefont {H.-Y.}\ \bibnamefont
  {Huang}}, \bibinfo {author} {\bibfnamefont {R.}~\bibnamefont {Kueng}},\ and\
  \bibinfo {author} {\bibfnamefont {J.}~\bibnamefont {Preskill}},\ }\bibfield
  {title} {\bibinfo {title} {Predicting many properties of a quantum system
  from very few measurements},\ }\href
  {https://doi.org/10.1038/s41567-020-0932-7} {\bibfield  {journal} {\bibinfo
  {journal} {Nat. Phys.}\ }\textbf {\bibinfo {volume} {16}},\ \bibinfo {pages}
  {1050} (\bibinfo {year} {2020})}\BibitemShut {NoStop}%
\bibitem [{\citenamefont {{\v{Z}}nidari{\v{c}}}\ \emph
  {et~al.}(2007)\citenamefont {{\v{Z}}nidari{\v{c}}}, \citenamefont {Prosen},
  \citenamefont {Benenti},\ and\ \citenamefont {Casati}}]{nidari2007}%
  \BibitemOpen
  \bibfield  {author} {\bibinfo {author} {\bibfnamefont {M.}~\bibnamefont
  {{\v{Z}}nidari{\v{c}}}}, \bibinfo {author} {\bibfnamefont {T.}~\bibnamefont
  {Prosen}}, \bibinfo {author} {\bibfnamefont {G.}~\bibnamefont {Benenti}},\
  and\ \bibinfo {author} {\bibfnamefont {G.}~\bibnamefont {Casati}},\
  }\bibfield  {title} {\bibinfo {title} {Detecting entanglement of random
  states with an entanglement witness},\ }\href
  {https://doi.org/10.1088/1751-8113/40/45/017} {\bibfield  {journal} {\bibinfo
   {journal} {J. Phys. A}\ }\textbf {\bibinfo {volume} {40}},\ \bibinfo {pages}
  {13787} (\bibinfo {year} {2007})}\BibitemShut {NoStop}%
\bibitem [{\citenamefont {Aubrun}\ \emph {et~al.}(2012)\citenamefont {Aubrun},
  \citenamefont {Szarek},\ and\ \citenamefont {Ye}}]{PhysRevA.85.030302}%
  \BibitemOpen
  \bibfield  {author} {\bibinfo {author} {\bibfnamefont {G.}~\bibnamefont
  {Aubrun}}, \bibinfo {author} {\bibfnamefont {S.~J.}\ \bibnamefont {Szarek}},\
  and\ \bibinfo {author} {\bibfnamefont {D.}~\bibnamefont {Ye}},\ }\bibfield
  {title} {\bibinfo {title} {Phase transitions for random states and a
  semicircle law for the partial transpose},\ }\href
  {https://doi.org/10.1103/PhysRevA.85.030302} {\bibfield  {journal} {\bibinfo
  {journal} {Phys. Rev. A}\ }\textbf {\bibinfo {volume} {85}},\ \bibinfo
  {pages} {030302} (\bibinfo {year} {2012})}\BibitemShut {NoStop}%
\bibitem [{\citenamefont {Bhosale}\ \emph {et~al.}(2012)\citenamefont
  {Bhosale}, \citenamefont {Tomsovic},\ and\ \citenamefont
  {Lakshminarayan}}]{PhysRevA.85.062331}%
  \BibitemOpen
  \bibfield  {author} {\bibinfo {author} {\bibfnamefont {U.~T.}\ \bibnamefont
  {Bhosale}}, \bibinfo {author} {\bibfnamefont {S.}~\bibnamefont {Tomsovic}},\
  and\ \bibinfo {author} {\bibfnamefont {A.}~\bibnamefont {Lakshminarayan}},\
  }\bibfield  {title} {\bibinfo {title} {Entanglement between two subsystems,
  the {Wigner} semicircle and extreme-value statistics},\ }\href
  {https://doi.org/10.1103/PhysRevA.85.062331} {\bibfield  {journal} {\bibinfo
  {journal} {Phys. Rev. A}\ }\textbf {\bibinfo {volume} {85}},\ \bibinfo
  {pages} {062331} (\bibinfo {year} {2012})}\BibitemShut {NoStop}%
\bibitem [{\citenamefont {Collins}\ and\ \citenamefont
  {Nechita}(2016)}]{Collins2016}%
  \BibitemOpen
  \bibfield  {author} {\bibinfo {author} {\bibfnamefont {B.}~\bibnamefont
  {Collins}}\ and\ \bibinfo {author} {\bibfnamefont {I.}~\bibnamefont
  {Nechita}},\ }\bibfield  {title} {\bibinfo {title} {Random matrix techniques
  in quantum information theory},\ }\href {https://doi.org/10.1063/1.4936880}
  {\bibfield  {journal} {\bibinfo  {journal} {J. Math. Phys.}\ }\textbf
  {\bibinfo {volume} {57}},\ \bibinfo {pages} {015215} (\bibinfo {year}
  {2016})}\BibitemShut {NoStop}%
\bibitem [{\citenamefont {Shapourian}\ \emph {et~al.}(2021)\citenamefont
  {Shapourian}, \citenamefont {Liu}, \citenamefont {Kudler-Flam},\ and\
  \citenamefont {Vishwanath}}]{PRXQuantum.2.030347}%
  \BibitemOpen
  \bibfield  {author} {\bibinfo {author} {\bibfnamefont {H.}~\bibnamefont
  {Shapourian}}, \bibinfo {author} {\bibfnamefont {S.}~\bibnamefont {Liu}},
  \bibinfo {author} {\bibfnamefont {J.}~\bibnamefont {Kudler-Flam}},\ and\
  \bibinfo {author} {\bibfnamefont {A.}~\bibnamefont {Vishwanath}},\ }\bibfield
   {title} {\bibinfo {title} {Entanglement negativity spectrum of random mixed
  states: A diagrammatic approach},\ }\href
  {https://doi.org/10.1103/PRXQuantum.2.030347} {\bibfield  {journal} {\bibinfo
   {journal} {PRX Quantum}\ }\textbf {\bibinfo {volume} {2}},\ \bibinfo {pages}
  {030347} (\bibinfo {year} {2021})}\BibitemShut {NoStop}%
\bibitem [{\citenamefont {Peres}(1996)}]{PhysRevLett.77.1413}%
  \BibitemOpen
  \bibfield  {author} {\bibinfo {author} {\bibfnamefont {A.}~\bibnamefont
  {Peres}},\ }\bibfield  {title} {\bibinfo {title} {Separability criterion for
  density matrices},\ }\href {https://doi.org/10.1103/PhysRevLett.77.1413}
  {\bibfield  {journal} {\bibinfo  {journal} {Phys. Rev. Lett.}\ }\textbf
  {\bibinfo {volume} {77}},\ \bibinfo {pages} {1413} (\bibinfo {year}
  {1996})}\BibitemShut {NoStop}%
\bibitem [{\citenamefont {Vidal}\ and\ \citenamefont
  {Werner}(2002)}]{PhysRevA.65.032314}%
  \BibitemOpen
  \bibfield  {author} {\bibinfo {author} {\bibfnamefont {G.}~\bibnamefont
  {Vidal}}\ and\ \bibinfo {author} {\bibfnamefont {R.~F.}\ \bibnamefont
  {Werner}},\ }\bibfield  {title} {\bibinfo {title} {Computable measure of
  entanglement},\ }\href {https://doi.org/10.1103/PhysRevA.65.032314}
  {\bibfield  {journal} {\bibinfo  {journal} {Phys. Rev. A}\ }\textbf {\bibinfo
  {volume} {65}},\ \bibinfo {pages} {032314} (\bibinfo {year}
  {2002})}\BibitemShut {NoStop}%
\bibitem [{\citenamefont {Plenio}(2005)}]{PhysRevLett.95.090503}%
  \BibitemOpen
  \bibfield  {author} {\bibinfo {author} {\bibfnamefont {M.~B.}\ \bibnamefont
  {Plenio}},\ }\bibfield  {title} {\bibinfo {title} {Logarithmic negativity: A
  full entanglement monotone that is not convex},\ }\href
  {https://doi.org/10.1103/PhysRevLett.95.090503} {\bibfield  {journal}
  {\bibinfo  {journal} {Phys. Rev. Lett.}\ }\textbf {\bibinfo {volume} {95}},\
  \bibinfo {pages} {090503} (\bibinfo {year} {2005})}\BibitemShut {NoStop}%
\bibitem [{\citenamefont {Carteret}(2005)}]{PhysRevLett.94.040502}%
  \BibitemOpen
  \bibfield  {author} {\bibinfo {author} {\bibfnamefont {H.~A.}\ \bibnamefont
  {Carteret}},\ }\bibfield  {title} {\bibinfo {title} {Noiseless quantum
  circuits for the {Peres} separability criterion},\ }\href
  {https://doi.org/10.1103/PhysRevLett.94.040502} {\bibfield  {journal}
  {\bibinfo  {journal} {Phys. Rev. Lett.}\ }\textbf {\bibinfo {volume} {94}},\
  \bibinfo {pages} {040502} (\bibinfo {year} {2005})}\BibitemShut {NoStop}%
\bibitem [{\citenamefont {Gray}\ \emph {et~al.}(2018)\citenamefont {Gray},
  \citenamefont {Banchi}, \citenamefont {Bayat},\ and\ \citenamefont
  {Bose}}]{PhysRevLett.121.150503}%
  \BibitemOpen
  \bibfield  {author} {\bibinfo {author} {\bibfnamefont {J.}~\bibnamefont
  {Gray}}, \bibinfo {author} {\bibfnamefont {L.}~\bibnamefont {Banchi}},
  \bibinfo {author} {\bibfnamefont {A.}~\bibnamefont {Bayat}},\ and\ \bibinfo
  {author} {\bibfnamefont {S.}~\bibnamefont {Bose}},\ }\bibfield  {title}
  {\bibinfo {title} {Machine-learning-assisted many-body entanglement
  measurement},\ }\href {https://doi.org/10.1103/PhysRevLett.121.150503}
  {\bibfield  {journal} {\bibinfo  {journal} {Phys. Rev. Lett.}\ }\textbf
  {\bibinfo {volume} {121}},\ \bibinfo {pages} {150503} (\bibinfo {year}
  {2018})}\BibitemShut {NoStop}%
\bibitem [{\citenamefont {Zhou}\ \emph {et~al.}(2020)\citenamefont {Zhou},
  \citenamefont {Zeng},\ and\ \citenamefont {Liu}}]{PhysRevLett.125.200502}%
  \BibitemOpen
  \bibfield  {author} {\bibinfo {author} {\bibfnamefont {Y.}~\bibnamefont
  {Zhou}}, \bibinfo {author} {\bibfnamefont {P.}~\bibnamefont {Zeng}},\ and\
  \bibinfo {author} {\bibfnamefont {Z.}~\bibnamefont {Liu}},\ }\bibfield
  {title} {\bibinfo {title} {Single-copies estimation of entanglement
  negativity},\ }\href {https://doi.org/10.1103/PhysRevLett.125.200502}
  {\bibfield  {journal} {\bibinfo  {journal} {Phys. Rev. Lett.}\ }\textbf
  {\bibinfo {volume} {125}},\ \bibinfo {pages} {200502} (\bibinfo {year}
  {2020})}\BibitemShut {NoStop}%
\bibitem [{\citenamefont {Ruggiero}\ \emph {et~al.}(2016)\citenamefont
  {Ruggiero}, \citenamefont {Alba},\ and\ \citenamefont
  {Calabrese}}]{PhysRevB.94.195121}%
  \BibitemOpen
  \bibfield  {author} {\bibinfo {author} {\bibfnamefont {P.}~\bibnamefont
  {Ruggiero}}, \bibinfo {author} {\bibfnamefont {V.}~\bibnamefont {Alba}},\
  and\ \bibinfo {author} {\bibfnamefont {P.}~\bibnamefont {Calabrese}},\
  }\bibfield  {title} {\bibinfo {title} {Negativity spectrum of one-dimensional
  conformal field theories},\ }\href
  {https://doi.org/10.1103/PhysRevB.94.195121} {\bibfield  {journal} {\bibinfo
  {journal} {Phys. Rev. B}\ }\textbf {\bibinfo {volume} {94}},\ \bibinfo
  {pages} {195121} (\bibinfo {year} {2016})}\BibitemShut {NoStop}%
\bibitem [{\citenamefont {Turkeshi}\ \emph {et~al.}(2020)\citenamefont
  {Turkeshi}, \citenamefont {Ruggiero},\ and\ \citenamefont
  {Calabrese}}]{PhysRevB.101.064207}%
  \BibitemOpen
  \bibfield  {author} {\bibinfo {author} {\bibfnamefont {X.}~\bibnamefont
  {Turkeshi}}, \bibinfo {author} {\bibfnamefont {P.}~\bibnamefont {Ruggiero}},\
  and\ \bibinfo {author} {\bibfnamefont {P.}~\bibnamefont {Calabrese}},\
  }\bibfield  {title} {\bibinfo {title} {Negativity spectrum in the random
  singlet phase},\ }\href {https://doi.org/10.1103/PhysRevB.101.064207}
  {\bibfield  {journal} {\bibinfo  {journal} {Phys. Rev. B}\ }\textbf {\bibinfo
  {volume} {101}},\ \bibinfo {pages} {064207} (\bibinfo {year}
  {2020})}\BibitemShut {NoStop}%
\bibitem [{\citenamefont {Murciano}\ \emph {et~al.}(2022)\citenamefont
  {Murciano}, \citenamefont {Vitale}, \citenamefont {Dalmonte},\ and\
  \citenamefont {Calabrese}}]{PhysRevLett.128.140502}%
  \BibitemOpen
  \bibfield  {author} {\bibinfo {author} {\bibfnamefont {S.}~\bibnamefont
  {Murciano}}, \bibinfo {author} {\bibfnamefont {V.}~\bibnamefont {Vitale}},
  \bibinfo {author} {\bibfnamefont {M.}~\bibnamefont {Dalmonte}},\ and\
  \bibinfo {author} {\bibfnamefont {P.}~\bibnamefont {Calabrese}},\ }\bibfield
  {title} {\bibinfo {title} {Negativity {Hamiltonian}: An operator
  characterization of mixed-state entanglement},\ }\href
  {https://doi.org/10.1103/PhysRevLett.128.140502} {\bibfield  {journal}
  {\bibinfo  {journal} {Phys. Rev. Lett.}\ }\textbf {\bibinfo {volume} {128}},\
  \bibinfo {pages} {140502} (\bibinfo {year} {2022})}\BibitemShut {NoStop}%
\bibitem [{\citenamefont {Elben}\ \emph {et~al.}(2020)\citenamefont {Elben},
  \citenamefont {Kueng}, \citenamefont {Huang}, \citenamefont {van Bijnen},
  \citenamefont {Kokail}, \citenamefont {Dalmonte}, \citenamefont {Calabrese},
  \citenamefont {Kraus}, \citenamefont {Preskill}, \citenamefont {Zoller},\
  and\ \citenamefont {Vermersch}}]{PhysRevLett.125.200501}%
  \BibitemOpen
  \bibfield  {author} {\bibinfo {author} {\bibfnamefont {A.}~\bibnamefont
  {Elben}}, \bibinfo {author} {\bibfnamefont {R.}~\bibnamefont {Kueng}},
  \bibinfo {author} {\bibfnamefont {H.-Y.~R.}\ \bibnamefont {Huang}}, \bibinfo
  {author} {\bibfnamefont {R.}~\bibnamefont {van Bijnen}}, \bibinfo {author}
  {\bibfnamefont {C.}~\bibnamefont {Kokail}}, \bibinfo {author} {\bibfnamefont
  {M.}~\bibnamefont {Dalmonte}}, \bibinfo {author} {\bibfnamefont
  {P.}~\bibnamefont {Calabrese}}, \bibinfo {author} {\bibfnamefont
  {B.}~\bibnamefont {Kraus}}, \bibinfo {author} {\bibfnamefont
  {J.}~\bibnamefont {Preskill}}, \bibinfo {author} {\bibfnamefont
  {P.}~\bibnamefont {Zoller}},\ and\ \bibinfo {author} {\bibfnamefont
  {B.}~\bibnamefont {Vermersch}},\ }\bibfield  {title} {\bibinfo {title}
  {Mixed-state entanglement from local randomized measurements},\ }\href
  {https://doi.org/10.1103/PhysRevLett.125.200501} {\bibfield  {journal}
  {\bibinfo  {journal} {Phys. Rev. Lett.}\ }\textbf {\bibinfo {volume} {125}},\
  \bibinfo {pages} {200501} (\bibinfo {year} {2020})}\BibitemShut {NoStop}%
\bibitem [{\citenamefont {Neven}\ \emph {et~al.}(2021)\citenamefont {Neven},
  \citenamefont {Carrasco}, \citenamefont {Vitale}, \citenamefont {Kokail},
  \citenamefont {Elben}, \citenamefont {Dalmonte}, \citenamefont {Calabrese},
  \citenamefont {Zoller}, \citenamefont {Vermersch}, \citenamefont {Kueng},\
  and\ \citenamefont {Kraus}}]{Neven2021}%
  \BibitemOpen
  \bibfield  {author} {\bibinfo {author} {\bibfnamefont {A.}~\bibnamefont
  {Neven}}, \bibinfo {author} {\bibfnamefont {J.}~\bibnamefont {Carrasco}},
  \bibinfo {author} {\bibfnamefont {V.}~\bibnamefont {Vitale}}, \bibinfo
  {author} {\bibfnamefont {C.}~\bibnamefont {Kokail}}, \bibinfo {author}
  {\bibfnamefont {A.}~\bibnamefont {Elben}}, \bibinfo {author} {\bibfnamefont
  {M.}~\bibnamefont {Dalmonte}}, \bibinfo {author} {\bibfnamefont
  {P.}~\bibnamefont {Calabrese}}, \bibinfo {author} {\bibfnamefont
  {P.}~\bibnamefont {Zoller}}, \bibinfo {author} {\bibfnamefont
  {B.}~\bibnamefont {Vermersch}}, \bibinfo {author} {\bibfnamefont
  {R.}~\bibnamefont {Kueng}},\ and\ \bibinfo {author} {\bibfnamefont
  {B.}~\bibnamefont {Kraus}},\ }\bibfield  {title} {\bibinfo {title}
  {Symmetry-resolved entanglement detection using partial transpose moments},\
  }\href {https://doi.org/10.1038/s41534-021-00487-y} {\bibfield  {journal}
  {\bibinfo  {journal} {npj Quantum Inf.}\ }\textbf {\bibinfo {volume} {7}},\
  \bibinfo {pages} {152} (\bibinfo {year} {2021})}\BibitemShut {NoStop}%
\bibitem [{sup()}]{supplementary}%
  \BibitemOpen
  \href@noop {} {}\bibinfo {note} {See supplementary material.}\BibitemShut
  {Stop}%
\bibitem [{\citenamefont {Pozniak}\ \emph {et~al.}(1998)\citenamefont
  {Pozniak}, \citenamefont {Zyczkowski},\ and\ \citenamefont
  {Kus}}]{Pozniak_1998}%
  \BibitemOpen
  \bibfield  {author} {\bibinfo {author} {\bibfnamefont {M.}~\bibnamefont
  {Pozniak}}, \bibinfo {author} {\bibfnamefont {K.}~\bibnamefont
  {Zyczkowski}},\ and\ \bibinfo {author} {\bibfnamefont {M.}~\bibnamefont
  {Kus}},\ }\bibfield  {title} {\bibinfo {title} {Composed ensembles of random
  unitary matrices},\ }\href {https://doi.org/10.1088/0305-4470/31/3/016}
  {\bibfield  {journal} {\bibinfo  {journal} {J. Phys. A}\ }\textbf {\bibinfo
  {volume} {31}},\ \bibinfo {pages} {1059} (\bibinfo {year}
  {1998})}\BibitemShut {NoStop}%
\bibitem [{\citenamefont {Emerson}\ \emph {et~al.}(2003)\citenamefont
  {Emerson}, \citenamefont {Weinstein}, \citenamefont {Saraceno}, \citenamefont
  {Lloyd},\ and\ \citenamefont {Cory}}]{science.1090790}%
  \BibitemOpen
  \bibfield  {author} {\bibinfo {author} {\bibfnamefont {J.}~\bibnamefont
  {Emerson}}, \bibinfo {author} {\bibfnamefont {Y.~S.}\ \bibnamefont
  {Weinstein}}, \bibinfo {author} {\bibfnamefont {M.}~\bibnamefont {Saraceno}},
  \bibinfo {author} {\bibfnamefont {S.}~\bibnamefont {Lloyd}},\ and\ \bibinfo
  {author} {\bibfnamefont {D.~G.}\ \bibnamefont {Cory}},\ }\bibfield  {title}
  {\bibinfo {title} {Pseudo-random unitary operators for quantum information
  processing},\ }\href {https://doi.org/10.1126/science.1090790} {\bibfield
  {journal} {\bibinfo  {journal} {Science}\ }\textbf {\bibinfo {volume}
  {302}},\ \bibinfo {pages} {2098} (\bibinfo {year} {2003})}\BibitemShut
  {NoStop}%
\bibitem [{\citenamefont {\ifmmode \check{Z}\else
  \v{Z}\fi{}nidari\ifmmode~\check{c}\else
  \v{c}\fi{}}(2007)}]{PhysRevA.76.012318}%
  \BibitemOpen
  \bibfield  {author} {\bibinfo {author} {\bibfnamefont {M.}~\bibnamefont
  {\ifmmode \check{Z}\else \v{Z}\fi{}nidari\ifmmode~\check{c}\else
  \v{c}\fi{}}},\ }\bibfield  {title} {\bibinfo {title} {Optimal two-qubit gate
  for generation of random bipartite entanglement},\ }\href
  {https://doi.org/10.1103/PhysRevA.76.012318} {\bibfield  {journal} {\bibinfo
  {journal} {Phys. Rev. A}\ }\textbf {\bibinfo {volume} {76}},\ \bibinfo
  {pages} {012318} (\bibinfo {year} {2007})}\BibitemShut {NoStop}%
\bibitem [{\citenamefont {Arnaud}\ and\ \citenamefont
  {Braun}(2008)}]{PhysRevA.78.062329}%
  \BibitemOpen
  \bibfield  {author} {\bibinfo {author} {\bibfnamefont {L.}~\bibnamefont
  {Arnaud}}\ and\ \bibinfo {author} {\bibfnamefont {D.}~\bibnamefont {Braun}},\
  }\bibfield  {title} {\bibinfo {title} {Efficiency of producing random unitary
  matrices with quantum circuits},\ }\href
  {https://doi.org/10.1103/PhysRevA.78.062329} {\bibfield  {journal} {\bibinfo
  {journal} {Phys. Rev. A}\ }\textbf {\bibinfo {volume} {78}},\ \bibinfo
  {pages} {062329} (\bibinfo {year} {2008})}\BibitemShut {NoStop}%
\bibitem [{\citenamefont {Harrow}\ and\ \citenamefont
  {Low}(2009)}]{Harrow2009}%
  \BibitemOpen
  \bibfield  {author} {\bibinfo {author} {\bibfnamefont {A.~W.}\ \bibnamefont
  {Harrow}}\ and\ \bibinfo {author} {\bibfnamefont {R.~A.}\ \bibnamefont
  {Low}},\ }\bibfield  {title} {\bibinfo {title} {Random quantum circuits are
  approximate 2-designs},\ }\href {https://doi.org/10.1007/s00220-009-0873-6}
  {\bibfield  {journal} {\bibinfo  {journal} {Commun. Math. Phys.}\ }\textbf
  {\bibinfo {volume} {291}},\ \bibinfo {pages} {257} (\bibinfo {year}
  {2009})}\BibitemShut {NoStop}%
\bibitem [{\citenamefont {Brown}\ and\ \citenamefont
  {Viola}(2010)}]{PhysRevLett.104.250501}%
  \BibitemOpen
  \bibfield  {author} {\bibinfo {author} {\bibfnamefont {W.~G.}\ \bibnamefont
  {Brown}}\ and\ \bibinfo {author} {\bibfnamefont {L.}~\bibnamefont {Viola}},\
  }\bibfield  {title} {\bibinfo {title} {Convergence rates for arbitrary
  statistical moments of random quantum circuits},\ }\href
  {https://doi.org/10.1103/PhysRevLett.104.250501} {\bibfield  {journal}
  {\bibinfo  {journal} {Phys. Rev. Lett.}\ }\textbf {\bibinfo {volume} {104}},\
  \bibinfo {pages} {250501} (\bibinfo {year} {2010})}\BibitemShut {NoStop}%
\bibitem [{\citenamefont {{\'{C}}wikli{\'{n}}ski}\ \emph
  {et~al.}(2013)\citenamefont {{\'{C}}wikli{\'{n}}ski}, \citenamefont
  {Horodecki}, \citenamefont {Mozrzymas}, \citenamefont {Pankowski},\ and\
  \citenamefont {Studzi{\'{n}}ski}}]{wikli_ski_2013}%
  \BibitemOpen
  \bibfield  {author} {\bibinfo {author} {\bibfnamefont {P.}~\bibnamefont
  {{\'{C}}wikli{\'{n}}ski}}, \bibinfo {author} {\bibfnamefont {M.}~\bibnamefont
  {Horodecki}}, \bibinfo {author} {\bibfnamefont {M.}~\bibnamefont
  {Mozrzymas}}, \bibinfo {author} {\bibfnamefont {{\L}.}~\bibnamefont
  {Pankowski}},\ and\ \bibinfo {author} {\bibfnamefont {M.}~\bibnamefont
  {Studzi{\'{n}}ski}},\ }\bibfield  {title} {\bibinfo {title} {Local random
  quantum circuits are approximate polynomial-designs: numerical results},\
  }\href {https://doi.org/10.1088/1751-8113/46/30/305301} {\bibfield  {journal}
  {\bibinfo  {journal} {J. Phys. A}\ }\textbf {\bibinfo {volume} {46}},\
  \bibinfo {pages} {305301} (\bibinfo {year} {2013})}\BibitemShut {NoStop}%
\bibitem [{\citenamefont {Brand\~ao}\ \emph {et~al.}(2016)\citenamefont
  {Brand\~ao}, \citenamefont {Harrow},\ and\ \citenamefont
  {Horodecki}}]{PhysRevLett.116.170502}%
  \BibitemOpen
  \bibfield  {author} {\bibinfo {author} {\bibfnamefont {F.~G. S.~L.}\
  \bibnamefont {Brand\~ao}}, \bibinfo {author} {\bibfnamefont {A.~W.}\
  \bibnamefont {Harrow}},\ and\ \bibinfo {author} {\bibfnamefont
  {M.}~\bibnamefont {Horodecki}},\ }\bibfield  {title} {\bibinfo {title}
  {Efficient quantum pseudorandomness},\ }\href
  {https://doi.org/10.1103/PhysRevLett.116.170502} {\bibfield  {journal}
  {\bibinfo  {journal} {Phys. Rev. Lett.}\ }\textbf {\bibinfo {volume} {116}},\
  \bibinfo {pages} {170502} (\bibinfo {year} {2016})}\BibitemShut {NoStop}%
\bibitem [{\citenamefont {Bensa}\ and\ \citenamefont {\ifmmode \check{Z}\else
  \v{Z}\fi{}nidari\ifmmode~\check{c}\else
  \v{c}\fi{}}(2021)}]{PhysRevX.11.031019}%
  \BibitemOpen
  \bibfield  {author} {\bibinfo {author} {\bibfnamefont {J.}~\bibnamefont
  {Bensa}}\ and\ \bibinfo {author} {\bibfnamefont {M.}~\bibnamefont {\ifmmode
  \check{Z}\else \v{Z}\fi{}nidari\ifmmode~\check{c}\else \v{c}\fi{}}},\
  }\bibfield  {title} {\bibinfo {title} {Fastest local entanglement scrambler,
  multistage thermalization, and a non-{Hermitian} phantom},\ }\href
  {https://doi.org/10.1103/PhysRevX.11.031019} {\bibfield  {journal} {\bibinfo
  {journal} {Phys. Rev. X}\ }\textbf {\bibinfo {volume} {11}},\ \bibinfo
  {pages} {031019} (\bibinfo {year} {2021})}\BibitemShut {NoStop}%
\bibitem [{\citenamefont {Turner}\ and\ \citenamefont
  {Markham}(2016)}]{PhysRevLett.116.200501}%
  \BibitemOpen
  \bibfield  {author} {\bibinfo {author} {\bibfnamefont {P.~S.}\ \bibnamefont
  {Turner}}\ and\ \bibinfo {author} {\bibfnamefont {D.}~\bibnamefont
  {Markham}},\ }\bibfield  {title} {\bibinfo {title} {Derandomizing quantum
  circuits with measurement-based unitary designs},\ }\href
  {https://doi.org/10.1103/PhysRevLett.116.200501} {\bibfield  {journal}
  {\bibinfo  {journal} {Phys. Rev. Lett.}\ }\textbf {\bibinfo {volume} {116}},\
  \bibinfo {pages} {200501} (\bibinfo {year} {2016})}\BibitemShut {NoStop}%
\bibitem [{\citenamefont {Mezher}\ \emph {et~al.}(2018)\citenamefont {Mezher},
  \citenamefont {Ghalbouni}, \citenamefont {Dgheim},\ and\ \citenamefont
  {Markham}}]{PhysRevA.97.022333}%
  \BibitemOpen
  \bibfield  {author} {\bibinfo {author} {\bibfnamefont {R.}~\bibnamefont
  {Mezher}}, \bibinfo {author} {\bibfnamefont {J.}~\bibnamefont {Ghalbouni}},
  \bibinfo {author} {\bibfnamefont {J.}~\bibnamefont {Dgheim}},\ and\ \bibinfo
  {author} {\bibfnamefont {D.}~\bibnamefont {Markham}},\ }\bibfield  {title}
  {\bibinfo {title} {Efficient quantum pseudorandomness with simple graph
  states},\ }\href {https://doi.org/10.1103/PhysRevA.97.022333} {\bibfield
  {journal} {\bibinfo  {journal} {Phys. Rev. A}\ }\textbf {\bibinfo {volume}
  {97}},\ \bibinfo {pages} {022333} (\bibinfo {year} {2018})}\BibitemShut
  {NoStop}%
\bibitem [{\citenamefont {Nakata}\ \emph {et~al.}(2017)\citenamefont {Nakata},
  \citenamefont {Hirche}, \citenamefont {Koashi},\ and\ \citenamefont
  {Winter}}]{PhysRevX.7.021006}%
  \BibitemOpen
  \bibfield  {author} {\bibinfo {author} {\bibfnamefont {Y.}~\bibnamefont
  {Nakata}}, \bibinfo {author} {\bibfnamefont {C.}~\bibnamefont {Hirche}},
  \bibinfo {author} {\bibfnamefont {M.}~\bibnamefont {Koashi}},\ and\ \bibinfo
  {author} {\bibfnamefont {A.}~\bibnamefont {Winter}},\ }\bibfield  {title}
  {\bibinfo {title} {Efficient quantum pseudorandomness with nearly
  time-independent hamiltonian dynamics},\ }\href
  {https://doi.org/10.1103/PhysRevX.7.021006} {\bibfield  {journal} {\bibinfo
  {journal} {Phys. Rev. X}\ }\textbf {\bibinfo {volume} {7}},\ \bibinfo {pages}
  {021006} (\bibinfo {year} {2017})}\BibitemShut {NoStop}%
\bibitem [{\citenamefont {Li}\ \emph {et~al.}(2019{\natexlab{a}})\citenamefont
  {Li}, \citenamefont {Luo}, \citenamefont {Xin}, \citenamefont {Wang},
  \citenamefont {Kribs}, \citenamefont {Lu}, \citenamefont {Zeng},\ and\
  \citenamefont {Laflamme}}]{PhysRevLett.123.030502}%
  \BibitemOpen
  \bibfield  {author} {\bibinfo {author} {\bibfnamefont {J.}~\bibnamefont
  {Li}}, \bibinfo {author} {\bibfnamefont {Z.}~\bibnamefont {Luo}}, \bibinfo
  {author} {\bibfnamefont {T.}~\bibnamefont {Xin}}, \bibinfo {author}
  {\bibfnamefont {H.}~\bibnamefont {Wang}}, \bibinfo {author} {\bibfnamefont
  {D.}~\bibnamefont {Kribs}}, \bibinfo {author} {\bibfnamefont
  {D.}~\bibnamefont {Lu}}, \bibinfo {author} {\bibfnamefont {B.}~\bibnamefont
  {Zeng}},\ and\ \bibinfo {author} {\bibfnamefont {R.}~\bibnamefont
  {Laflamme}},\ }\bibfield  {title} {\bibinfo {title} {Experimental
  implementation of efficient quantum pseudorandomness on a 12-spin system},\
  }\href {https://doi.org/10.1103/PhysRevLett.123.030502} {\bibfield  {journal}
  {\bibinfo  {journal} {Phys. Rev. Lett.}\ }\textbf {\bibinfo {volume} {123}},\
  \bibinfo {pages} {030502} (\bibinfo {year} {2019}{\natexlab{a}})}\BibitemShut
  {NoStop}%
\bibitem [{\citenamefont {Song}\ \emph {et~al.}(2017)\citenamefont {Song},
  \citenamefont {Xu}, \citenamefont {Liu}, \citenamefont {Yang}, \citenamefont
  {Zheng}, \citenamefont {Deng}, \citenamefont {Xie}, \citenamefont {Huang},
  \citenamefont {Guo}, \citenamefont {Zhang}, \citenamefont {Zhang},
  \citenamefont {Xu}, \citenamefont {Zheng}, \citenamefont {Zhu}, \citenamefont
  {Wang}, \citenamefont {Chen}, \citenamefont {Lu}, \citenamefont {Han},\ and\
  \citenamefont {Pan}}]{PhysRevLett.119.180511}%
  \BibitemOpen
  \bibfield  {author} {\bibinfo {author} {\bibfnamefont {C.}~\bibnamefont
  {Song}}, \bibinfo {author} {\bibfnamefont {K.}~\bibnamefont {Xu}}, \bibinfo
  {author} {\bibfnamefont {W.}~\bibnamefont {Liu}}, \bibinfo {author}
  {\bibfnamefont {C.-p.}\ \bibnamefont {Yang}}, \bibinfo {author}
  {\bibfnamefont {S.-B.}\ \bibnamefont {Zheng}}, \bibinfo {author}
  {\bibfnamefont {H.}~\bibnamefont {Deng}}, \bibinfo {author} {\bibfnamefont
  {Q.}~\bibnamefont {Xie}}, \bibinfo {author} {\bibfnamefont {K.}~\bibnamefont
  {Huang}}, \bibinfo {author} {\bibfnamefont {Q.}~\bibnamefont {Guo}}, \bibinfo
  {author} {\bibfnamefont {L.}~\bibnamefont {Zhang}}, \bibinfo {author}
  {\bibfnamefont {P.}~\bibnamefont {Zhang}}, \bibinfo {author} {\bibfnamefont
  {D.}~\bibnamefont {Xu}}, \bibinfo {author} {\bibfnamefont {D.}~\bibnamefont
  {Zheng}}, \bibinfo {author} {\bibfnamefont {X.}~\bibnamefont {Zhu}}, \bibinfo
  {author} {\bibfnamefont {H.}~\bibnamefont {Wang}}, \bibinfo {author}
  {\bibfnamefont {Y.-A.}\ \bibnamefont {Chen}}, \bibinfo {author}
  {\bibfnamefont {C.-Y.}\ \bibnamefont {Lu}}, \bibinfo {author} {\bibfnamefont
  {S.}~\bibnamefont {Han}},\ and\ \bibinfo {author} {\bibfnamefont {J.-W.}\
  \bibnamefont {Pan}},\ }\bibfield  {title} {\bibinfo {title} {10-qubit
  entanglement and parallel logic operations with a superconducting circuit},\
  }\href {https://doi.org/10.1103/PhysRevLett.119.180511} {\bibfield  {journal}
  {\bibinfo  {journal} {Phys. Rev. Lett.}\ }\textbf {\bibinfo {volume} {119}},\
  \bibinfo {pages} {180511} (\bibinfo {year} {2017})}\BibitemShut {NoStop}%
\bibitem [{\citenamefont {Song}\ \emph {et~al.}(2019)\citenamefont {Song},
  \citenamefont {Xu}, \citenamefont {Li}, \citenamefont {Zhang}, \citenamefont
  {Zhang}, \citenamefont {Liu}, \citenamefont {Guo}, \citenamefont {Wang},
  \citenamefont {Ren}, \citenamefont {Hao}, \citenamefont {Feng}, \citenamefont
  {Fan}, \citenamefont {Zheng}, \citenamefont {Wang}, \citenamefont {Wang},\
  and\ \citenamefont {Zhu}}]{Schrodingercat20}%
  \BibitemOpen
  \bibfield  {author} {\bibinfo {author} {\bibfnamefont {C.}~\bibnamefont
  {Song}}, \bibinfo {author} {\bibfnamefont {K.}~\bibnamefont {Xu}}, \bibinfo
  {author} {\bibfnamefont {H.}~\bibnamefont {Li}}, \bibinfo {author}
  {\bibfnamefont {Y.-R.}\ \bibnamefont {Zhang}}, \bibinfo {author}
  {\bibfnamefont {X.}~\bibnamefont {Zhang}}, \bibinfo {author} {\bibfnamefont
  {W.}~\bibnamefont {Liu}}, \bibinfo {author} {\bibfnamefont {Q.}~\bibnamefont
  {Guo}}, \bibinfo {author} {\bibfnamefont {Z.}~\bibnamefont {Wang}}, \bibinfo
  {author} {\bibfnamefont {W.}~\bibnamefont {Ren}}, \bibinfo {author}
  {\bibfnamefont {J.}~\bibnamefont {Hao}}, \bibinfo {author} {\bibfnamefont
  {H.}~\bibnamefont {Feng}}, \bibinfo {author} {\bibfnamefont {H.}~\bibnamefont
  {Fan}}, \bibinfo {author} {\bibfnamefont {D.}~\bibnamefont {Zheng}}, \bibinfo
  {author} {\bibfnamefont {D.-W.}\ \bibnamefont {Wang}}, \bibinfo {author}
  {\bibfnamefont {H.}~\bibnamefont {Wang}},\ and\ \bibinfo {author}
  {\bibfnamefont {S.-Y.}\ \bibnamefont {Zhu}},\ }\bibfield  {title} {\bibinfo
  {title} {Generation of multicomponent atomic {Schr\"{o}dinger} cat states of
  up to 20 qubits},\ }\href {https://doi.org/10.1126/science.aay0600}
  {\bibfield  {journal} {\bibinfo  {journal} {Science}\ }\textbf {\bibinfo
  {volume} {365}},\ \bibinfo {pages} {574} (\bibinfo {year}
  {2019})}\BibitemShut {NoStop}%
\bibitem [{\citenamefont {Li}\ \emph {et~al.}(2019{\natexlab{b}})\citenamefont
  {Li}, \citenamefont {Dong}, \citenamefont {Song},\ and\ \citenamefont
  {Wang}}]{PhysRevA.100.062302}%
  \BibitemOpen
  \bibfield  {author} {\bibinfo {author} {\bibfnamefont {K.-M.}\ \bibnamefont
  {Li}}, \bibinfo {author} {\bibfnamefont {H.}~\bibnamefont {Dong}}, \bibinfo
  {author} {\bibfnamefont {C.}~\bibnamefont {Song}},\ and\ \bibinfo {author}
  {\bibfnamefont {H.}~\bibnamefont {Wang}},\ }\bibfield  {title} {\bibinfo
  {title} {Approaching the chaotic regime with a fully connected
  superconducting quantum processor},\ }\href
  {https://doi.org/10.1103/PhysRevA.100.062302} {\bibfield  {journal} {\bibinfo
   {journal} {Phys. Rev. A}\ }\textbf {\bibinfo {volume} {100}},\ \bibinfo
  {pages} {062302} (\bibinfo {year} {2019}{\natexlab{b}})}\BibitemShut
  {NoStop}%
\bibitem [{\citenamefont {{Shim}}\ \emph {et~al.}(2013)\citenamefont {{Shim}},
  \citenamefont {{Fei}}, \citenamefont {{Oh}}, \citenamefont {{Hu}},\ and\
  \citenamefont {{Friesen}}}]{arXiv1303.0297S}%
  \BibitemOpen
  \bibfield  {author} {\bibinfo {author} {\bibfnamefont {Y.-P.}\ \bibnamefont
  {{Shim}}}, \bibinfo {author} {\bibfnamefont {J.}~\bibnamefont {{Fei}}},
  \bibinfo {author} {\bibfnamefont {S.}~\bibnamefont {{Oh}}}, \bibinfo {author}
  {\bibfnamefont {X.}~\bibnamefont {{Hu}}},\ and\ \bibinfo {author}
  {\bibfnamefont {M.}~\bibnamefont {{Friesen}}},\ }\bibfield  {title} {\bibinfo
  {title} {{Single-qubit gates in two steps with rotation axes in a single
  plane}},\ }\href {https://doi.org/10.48550/arXiv.1303.0297} {\bibfield
  {journal} {\bibinfo  {journal} {arXiv}\ ,\ \bibinfo {pages} {1303.0297}}
  (\bibinfo {year} {2013})}\BibitemShut {NoStop}%
\bibitem [{\citenamefont {Emerson}\ \emph {et~al.}(2005)\citenamefont
  {Emerson}, \citenamefont {Livine},\ and\ \citenamefont
  {Lloyd}}]{PhysRevA.72.060302}%
  \BibitemOpen
  \bibfield  {author} {\bibinfo {author} {\bibfnamefont {J.}~\bibnamefont
  {Emerson}}, \bibinfo {author} {\bibfnamefont {E.}~\bibnamefont {Livine}},\
  and\ \bibinfo {author} {\bibfnamefont {S.}~\bibnamefont {Lloyd}},\ }\bibfield
   {title} {\bibinfo {title} {Convergence conditions for random quantum
  circuits},\ }\href {https://doi.org/10.1103/PhysRevA.72.060302} {\bibfield
  {journal} {\bibinfo  {journal} {Phys. Rev. A}\ }\textbf {\bibinfo {volume}
  {72}},\ \bibinfo {pages} {060302} (\bibinfo {year} {2005})}\BibitemShut
  {NoStop}%
\bibitem [{\citenamefont {Porter}\ and\ \citenamefont
  {Thomas}(1956)}]{PhysRev.104.483}%
  \BibitemOpen
  \bibfield  {author} {\bibinfo {author} {\bibfnamefont {C.~E.}\ \bibnamefont
  {Porter}}\ and\ \bibinfo {author} {\bibfnamefont {R.~G.}\ \bibnamefont
  {Thomas}},\ }\bibfield  {title} {\bibinfo {title} {Fluctuations of nuclear
  reaction widths},\ }\href {https://doi.org/10.1103/PhysRev.104.483}
  {\bibfield  {journal} {\bibinfo  {journal} {Phys. Rev.}\ }\textbf {\bibinfo
  {volume} {104}},\ \bibinfo {pages} {483} (\bibinfo {year}
  {1956})}\BibitemShut {NoStop}%
\bibitem [{\citenamefont {Neill}\ \emph {et~al.}(2018)\citenamefont {Neill},
  \citenamefont {Roushan}, \citenamefont {Kechedzhi}, \citenamefont {Boixo},
  \citenamefont {Isakov}, \citenamefont {Smelyanskiy}, \citenamefont {Megrant},
  \citenamefont {Chiaro}, \citenamefont {Dunsworth}, \citenamefont {Arya} \emph
  {et~al.}}]{Neill2018}%
  \BibitemOpen
  \bibfield  {author} {\bibinfo {author} {\bibfnamefont {C.}~\bibnamefont
  {Neill}}, \bibinfo {author} {\bibfnamefont {P.}~\bibnamefont {Roushan}},
  \bibinfo {author} {\bibfnamefont {K.}~\bibnamefont {Kechedzhi}}, \bibinfo
  {author} {\bibfnamefont {S.}~\bibnamefont {Boixo}}, \bibinfo {author}
  {\bibfnamefont {S.~V.}\ \bibnamefont {Isakov}}, \bibinfo {author}
  {\bibfnamefont {V.}~\bibnamefont {Smelyanskiy}}, \bibinfo {author}
  {\bibfnamefont {A.}~\bibnamefont {Megrant}}, \bibinfo {author} {\bibfnamefont
  {B.}~\bibnamefont {Chiaro}}, \bibinfo {author} {\bibfnamefont
  {A.}~\bibnamefont {Dunsworth}}, \bibinfo {author} {\bibfnamefont
  {K.}~\bibnamefont {Arya}}, \emph {et~al.},\ }\bibfield  {title} {\bibinfo
  {title} {A blueprint for demonstrating quantum supremacy with superconducting
  qubits},\ }\href {https://doi.org/10.1126/science.aao4309} {\bibfield
  {journal} {\bibinfo  {journal} {Science}\ }\textbf {\bibinfo {volume}
  {360}},\ \bibinfo {pages} {195} (\bibinfo {year} {2018})}\BibitemShut
  {NoStop}%
\bibitem [{\citenamefont {Wu}\ \emph {et~al.}(2021)\citenamefont {Wu},
  \citenamefont {Bao}, \citenamefont {Cao}, \citenamefont {Chen}, \citenamefont
  {Chen}, \citenamefont {Chen}, \citenamefont {Chung}, \citenamefont {Deng},
  \citenamefont {Du}, \citenamefont {Fan} \emph
  {et~al.}}]{PhysRevLett.127.180501}%
  \BibitemOpen
  \bibfield  {author} {\bibinfo {author} {\bibfnamefont {Y.}~\bibnamefont
  {Wu}}, \bibinfo {author} {\bibfnamefont {W.-S.}\ \bibnamefont {Bao}},
  \bibinfo {author} {\bibfnamefont {S.}~\bibnamefont {Cao}}, \bibinfo {author}
  {\bibfnamefont {F.}~\bibnamefont {Chen}}, \bibinfo {author} {\bibfnamefont
  {M.-C.}\ \bibnamefont {Chen}}, \bibinfo {author} {\bibfnamefont
  {X.}~\bibnamefont {Chen}}, \bibinfo {author} {\bibfnamefont {T.-H.}\
  \bibnamefont {Chung}}, \bibinfo {author} {\bibfnamefont {H.}~\bibnamefont
  {Deng}}, \bibinfo {author} {\bibfnamefont {Y.}~\bibnamefont {Du}}, \bibinfo
  {author} {\bibfnamefont {D.}~\bibnamefont {Fan}}, \emph {et~al.},\ }\bibfield
   {title} {\bibinfo {title} {Strong quantum computational advantage using a
  superconducting quantum processor},\ }\href
  {https://doi.org/10.1103/PhysRevLett.127.180501} {\bibfield  {journal}
  {\bibinfo  {journal} {Phys. Rev. Lett.}\ }\textbf {\bibinfo {volume} {127}},\
  \bibinfo {pages} {180501} (\bibinfo {year} {2021})}\BibitemShut {NoStop}%
\bibitem [{\citenamefont {Nahum}\ \emph {et~al.}(2017)\citenamefont {Nahum},
  \citenamefont {Ruhman}, \citenamefont {Vijay},\ and\ \citenamefont
  {Haah}}]{PhysRevX.7.031016}%
  \BibitemOpen
  \bibfield  {author} {\bibinfo {author} {\bibfnamefont {A.}~\bibnamefont
  {Nahum}}, \bibinfo {author} {\bibfnamefont {J.}~\bibnamefont {Ruhman}},
  \bibinfo {author} {\bibfnamefont {S.}~\bibnamefont {Vijay}},\ and\ \bibinfo
  {author} {\bibfnamefont {J.}~\bibnamefont {Haah}},\ }\bibfield  {title}
  {\bibinfo {title} {Quantum entanglement growth under random unitary
  dynamics},\ }\href {https://doi.org/10.1103/PhysRevX.7.031016} {\bibfield
  {journal} {\bibinfo  {journal} {Phys. Rev. X}\ }\textbf {\bibinfo {volume}
  {7}},\ \bibinfo {pages} {031016} (\bibinfo {year} {2017})}\BibitemShut
  {NoStop}%
\bibitem [{\citenamefont {Nahum}\ \emph {et~al.}(2018)\citenamefont {Nahum},
  \citenamefont {Vijay},\ and\ \citenamefont {Haah}}]{PhysRevX.8.021014}%
  \BibitemOpen
  \bibfield  {author} {\bibinfo {author} {\bibfnamefont {A.}~\bibnamefont
  {Nahum}}, \bibinfo {author} {\bibfnamefont {S.}~\bibnamefont {Vijay}},\ and\
  \bibinfo {author} {\bibfnamefont {J.}~\bibnamefont {Haah}},\ }\bibfield
  {title} {\bibinfo {title} {Operator spreading in random unitary circuits},\
  }\href {https://doi.org/10.1103/PhysRevX.8.021014} {\bibfield  {journal}
  {\bibinfo  {journal} {Phys. Rev. X}\ }\textbf {\bibinfo {volume} {8}},\
  \bibinfo {pages} {021014} (\bibinfo {year} {2018})}\BibitemShut {NoStop}%
\bibitem [{\citenamefont {Lu}\ and\ \citenamefont
  {Grover}(2020)}]{PhysRevB.102.235110}%
  \BibitemOpen
  \bibfield  {author} {\bibinfo {author} {\bibfnamefont {T.-C.}\ \bibnamefont
  {Lu}}\ and\ \bibinfo {author} {\bibfnamefont {T.}~\bibnamefont {Grover}},\
  }\bibfield  {title} {\bibinfo {title} {Entanglement transitions as a probe of
  quasiparticles and quantum thermalization},\ }\href
  {https://doi.org/10.1103/PhysRevB.102.235110} {\bibfield  {journal} {\bibinfo
   {journal} {Phys. Rev. B}\ }\textbf {\bibinfo {volume} {102}},\ \bibinfo
  {pages} {235110} (\bibinfo {year} {2020})}\BibitemShut {NoStop}%
\end{thebibliography}
\end{document}